\documentclass[sigconf]{acmart}
\makeatletter
\def\@ACM@checkaffil{
    \if@ACM@instpresent\else
    \ClassWarningNoLine{\@classname}{No institution present for an affiliation}%
    \fi
    \if@ACM@citypresent\else
    \ClassWarningNoLine{\@classname}{No city present for an affiliation}%
    \fi
    \if@ACM@countrypresent\else
        \ClassWarningNoLine{\@classname}{No country present for an affiliation}%
    \fi
}
\makeatother
\usepackage[T1]{fontenc}
\usepackage{color}
\usepackage{balance}
\usepackage{booktabs}
\usepackage{textcomp}
\usepackage{stmaryrd}
\usepackage{graphicx}
\usepackage[caption=false,font=footnotesize]{subfig}
\usepackage{hyperref}
\hypersetup{colorlinks=true}
\definecolor{Salmon}{rgb}{1.0, 0.57, 0.64}
\definecolor{Cyan}{rgb}{0.0, 1.0, 1.0}
\definecolor{lgray}{rgb}{0.95, 0.95, 0.95}
\usepackage{mathtools}
\usepackage{supertabular,booktabs}
\usepackage{longtable}
\makeatother

\usepackage[frozencache,cachedir=_minted-bragghls]{minted}

\renewcommand{\listingscaption}{Listing}

\definecolor{bg}{rgb}{0.95,0.95,0.95}
\setminted{
    bgcolor=bg,
    escapeinside={||},
    mathescape=true,
}
\setmintedinline{breaklines}
\AtBeginEnvironment{minted}{%
  \catcode`?\active
  \begingroup\lccode`~=`\?\lowercase{\endgroup\def~{\linebreak}}%
}
\DeclareRobustCommand{\perc}{{%
  \mbox{%
    \fontencoding{\encodingdefault}%
    \fontfamily{pcr}%
    \selectfont
    \symbol{`\%}%
  }%
}}

\begin{document}

\title[\texttt{OpenHLS}]{\texttt{OpenHLS}: High-Level Synthesis for Low-Latency Deep Neural Networks for Experimental Science}
\author{Maksim Levental}
\email{mlevental@uchicago.edu}
\affiliation{%
  \institution{University of Chicago}
}
\author{Arham Khan}
\email{arham@uchicago.edu}
\affiliation{%
  \institution{University of Chicago}
}

\author{Ryan Chard}
\email{rchard@anl.gov}
\affiliation{%
  \institution{Argonne National Laboratory}
}
\author{Kazutomo Yoshii}
\email{kazutomo@anl.gov}
\affiliation{%
  \institution{Argonne National Laboratory}
}
\author{Kyle Chard}
\email{chard@uchicago.edu}
\affiliation{%
  \institution{University of Chicago}
}
\author{Ian Foster}
\email{foster@uchicago.edu}
\affiliation{%
  \institution{University of Chicago}
}

\begin{abstract}
In many experiment-driven scientific domains, such as high-energy
physics, material science, and cosmology, high data rate experiments
impose hard constraints on data acquisition systems:
collected data must either be indiscriminately stored for post-processing
and analysis, thereby necessitating large storage capacity, or accurately
filtered in real-time, thereby necessitating low-latency processing.
Deep neural networks, effective in other filtering tasks, have
not been widely employed in such data acquisition systems, due to
design and deployment difficulties. We present an open source,
lightweight, compiler framework, without any proprietary dependencies, \texttt{OpenHLS}, based on high-level
synthesis techniques, for translating high-level representations of
deep neural networks to low-level representations, suitable for deployment
to near-sensor devices such as field-programmable gate arrays. We
evaluate \texttt{OpenHLS} on various workloads and present a case-study
implementation of a deep neural network for Bragg peak detection in
the context of high-energy diffraction microscopy. We show \texttt{OpenHLS}
is able to produce an implementation of the network with a throughput
4.8 \textmu s/sample, which is approximately a 4$\times$ improvement
over the existing implementation.

\end{abstract}

\maketitle

\section{Introduction\label{sec:Introduction}}
High data rates are observed and, consequently, large datasets
are generated, across a broad range of science experiments in domains
such as high-energy physics, materials science, and cosmology. For
example, in high-energy physics, the LHCb detector at the Large
Hadron Collider (LHC) is tasked with observing the trajectories of particles
produced in proton-proton collisions at 40 MHz~\cite{pmlr-v42-glig14}. With a packet size of approximately
50 kB (per collision), this implies a data rate of approximately 2
TB/s. Ultimately, in combination with other detectors, the LHC processes
approximately 100 EB of data per year. In materials science, Bragg diffraction peak analysis, which provides non-destructive
characterization of single-crystal and polycrystalline structure and its evolution in a broad class of materials, can have collection
rates approaching 1 MHz~\cite{Hammer_2021}, with a corresponding
packet size of 80 kB. In cosmology, the Square Kilometer Array, a
radio telescope projected to be 
operational by 2027~\cite{mcmullin2022square}, will sustain data rates in excess
of 10 TB/s~\cite{grainge2017square}.

Storing and distributing
such large quantities of data 
for further analysis is cost prohibitive. 
Thus, data must be compressed or (as we consider here) filtered to preserve only the most ``interesting'' elements at the time of collection,
an approach that reduces storage needs but imposes
stringent latency constraints on the
filtering mechanisms. Typically, filtering mechanisms consist
of either physics-based~\cite{LHCB-FIGURE-2020-018} or machine
learning models~\cite{Gligorov_2013}; in either case, maximally efficient
and effective use of the target hardware platform is important. 
Irrespective
of the technique employed, almost universally, for ultra-low (e.g., sub-microsecond)
latency use cases the
implementation is deployed to either field-programmable gate arrays
(FPGAs) or application-specific integrated circuits (ASICs)~\cite{Duarte_2018}.
Here we focus primarily on FPGAs.

Deep neural networks (DNNs), a particular type of machine learning
model, have been shown to be effective in many scientific and commercial
domains due to their representational capacity, i.e., their ability to represent (approximately) diverse sets of mappings~\cite{alzubaidi2021review}.
DNNs ``learn'' to represent a mapping over the course of ``training,''
wherein they are iteratively evaluated on sample data while a ``learning
rule'' periodically updates the \emph{weights} that parameterize
the DNN. In recent years, DNNs have been investigated for near real-time
scientific use cases~\cite{liu2019deep,patton2018167,liu2022exploring}
but their use for the lowest latency use cases has been limited~\cite{Duarte_2018},
for three reasons:
\begin{enumerate}
\item Graphics Processing Units (GPUs), the conventional hardware target
for DNNs, are not sufficiently performant for
these high data rate, low latency, use cases (due to their
low clock speeds and low peripheral bandwidth, until recently~\cite{aaij2020allen});
\item DNNs, by virtue of their depth, require substantial memory (for weights) and compute (floating-point arithmetic),
thereby preventing their deployment to FPGAs, which, in particular,
have limited static RAM;
\item DNNs are (typically) defined, trained, and distributed by using high-level
frameworks (e.g., PyTorch~\cite{paszke2017automatic}, TensorFlow
\cite{https://doi.org/10.48550/arxiv.1603.04467}, MXNet~\cite{https://doi.org/10.48550/arxiv.1512.01274}),
which abstract all implementation details, thereby making
portability of model architectures to unsupported hardware platforms (e.g., FPGAs and ASICs) close to non-existent (barring almost wholesale reimplementations of the frameworks).
\end{enumerate}

These three barriers demand a solution that can translate a high-level DNN representation to a low-level representation,
suitable for FPGA deployment, while simultaneously optimizing resource usage and
minimizing latency. In general, the task of \emph{lowering} high-level
representations of programs to low-level representations is the domain
of a compiler. Similarly, the task of \emph{synthesizing} a \emph{register-transfer level} (RTL) \emph{design}, rendered in a \emph{hardware
description language} (HDL), from a program, is the domain of high-level
synthesis (HLS)~\cite{7368920} tools. Existing HLS tools~\cite{10.1145/2514740,Zhang2008,ferrandi2021bambu} struggle
to perform needed optimizations in reasonable amounts
of time (see Section~\ref{subsec:High-level-synthesis}) despite,
often, bundling robust optimizing compilers.

Recently, deep learning compilers (e.g., TVM~\cite{chen2018tvm},
MLIR~\cite{https://doi.org/10.48550/arxiv.2002.11054}, and Glow~\cite{https://doi.org/10.48550/arxiv.1805.00907})
have demonstrated the ability to reduce dramatically inference DNN latencies~\cite{https://doi.org/10.48550/arxiv.1809.02697}, training times~\cite{9664259}, and memory usage~\cite{https://doi.org/10.48550/arxiv.1604.06174}. These compilers function by extracting intermediate-level
representations (IRs) of the DNNs from the representations produced
by the frameworks, and performing various optimizations
(e.g., kernel fusion~\cite{10.1145/2858788.2688521}, vectorization~\cite{maleki2011evaluation}, and memory planning~\cite{https://doi.org/10.48550/arxiv.1604.06174}) on those IRs.
The highly optimized IR is then used to generate code for various
target hardware platforms. Given the successes of these compilers,
it is natural to wonder whether they can be adapted to the task of
sufficiently optimizing a DNN such that it might be synthesized to
RTL, for deployment to FPGA.

In this paper, we present \texttt{OpenHLS}, an open-source\footnote{Available
at \url{https://github.com/makslevental/openhls}}, lightweight
compiler and HLS framework that can translate DNNs defined as PyTorch
models to FPGA-compatible implementations. \texttt{OpenHLS} uses
a combination of compiler and HLS techniques to compile the entire
DNN into fully scheduled RTL, thereby eliminating all synchronization
overheads and achieving low latency. \texttt{OpenHLS} is general
and supports a wide range of DNN layer types, and thus a wide range of DNNs.
To the best of our knowledge, \texttt{OpenHLS} is the first HLS framework that enables the use of DNNs, free of a dependence on expensive and opaque proprietary HLS tools, for science experiments that demand low-latency inference.
In summary our
specific contributions include:
\begin{enumerate}
\item We describe and implement a compiler framework, \texttt{OpenHLS}, that can efficiently transform, without use of proprietary HLS tools, unoptimized, hardware-agnostic PyTorch models into low-latency RTL suitable for deployment to FPGAs;
\item We show that \texttt{OpenHLS} generates lower latency designs than does a state-of-the-art commercial HLS tool (Xilinx's Vitis
HLS) for many DNN layer types. In particular we show that
\texttt{OpenHLS} can produce synthesizable designs that meet placement,
routing, and timing constraints for \texttt{BraggNN}, a DNN designed for analyzing Bragg diffraction peaks;
\item We discuss challenges faced even after successful synthesis
of RTL from a high-level representation of a DNN, namely during the
place and route phases of implementation.
\end{enumerate}
Note that while we focus here, for illustrative purposes, on  optimizations relevant to a DNN used for identifying Bragg diffraction peaks in materials science,
\texttt{OpenHLS} supports a wide range of DNNs, limited only by upstream support for DNN layers.

The rest of this paper is as follows: Section~\ref{sec:Background}
reviews key concepts from compilers, high-level synthesis, and RTL
design for FPGA, as well as related work. Section~\ref{sec:OpenHLS-compiler-and} describes
the \texttt{OpenHLS} compiler and HLS framework in detail. Section
\ref{sec:Evaluation} evaluates \texttt{OpenHLS}\textquoteright s
performance, scalability, and competitiveness with designs generated
by Vitis HLS, and  
describes a
case study in which \texttt{OpenHLS} is applied to \texttt{BraggNN}, a
Bragg peak detection DNN with a target latency of 1 \textmu s/sample.
Finally, Section~\ref{sec:Conclusion} concludes and discusses future work.

\section{Background\label{sec:Background}}

We briefly review relevant concepts from DNN frameworks and compilers, high-level synthesis, and FPGA design. Each subsection corresponds to a phase in the translation from high-level DNN to feasible FPGA implementation.

\subsection{Compilers: The path from high to low}

The path from a high-level, abstract, DNN representation to a
register-transfer level representation can be viewed as a sequence
of 
lowerings between adjacent levels of abstraction. Each
level of abstraction is rendered as a programming language, IR, or
HDL, and thus we describe each lowering in terms of the representations
and tools used by \texttt{OpenHLS} to manipulate those representations:
\begin{enumerate}
\item An imperative, \emph{define-by-run,} Python representation, in PyTorch;
\item High-level data-flow graph representation, in TorchScript;
\item Low-level data and control flow graph representation, in Multi-Level Intermediate Representation (MLIR).
\end{enumerate}

\subsubsection{PyTorch and TorchScript}

Typically DNN models are represented in terms of high-level frameworks,
themselves implemented within general purpose programming languages.
Such frameworks are popular because of their ease of use and large
library of example implementations of various DNN model architectures.
\texttt{OpenHLS} targets the PyTorch framework.
DNNs developed within PyTorch are \emph{defined-by-run}:
the author describes the DNN imperatively in terms of high-level operations,
using Python, which, when executed, materializes the (partial) high-level
data-flow graph (DFG) corresponding to the DNN (e.g., for the purposes
of reverse-mode automatic differentiation). From the perspective of
the user, define-by-run enables fast iteration at development time,
possibly at the cost of some runtime performance.

Yet from the perspective of compilation, define-by-run
precludes efficient extraction of the high-level DFG; since the DFG
is materialized only at runtime, it cannot easily be statically inferred from
the textual representation (i.e., the Python source) of the DNN. Furthermore,
a priori, the runtime-materialized DFG is only partially materialized~\cite{paszke2017automatic},
and only as an in-memory data structure. Thus, framework support
is necessary for efficiently extracting the full DFG. For this purpose, PyTorch
supports a Single Static Assignment (SSA) IR, called TorchScript (TS)
IR and accompanying tracing mechanism (the TS JIT), which generates
TS IR from conventionally defined PyTorch models. Lowering from PyTorch
to TS IR enables various useful analyses and transformations on a
DNN at the level of the high-level DFG, but targeting FPGAs requires
a broader collection of transformations. To this end, we turn to a
recent addition to the compiler ecosystem, MLIR.

\subsubsection{MLIR\label{subsec:MLIR}}

MLIR~\cite{https://doi.org/10.48550/arxiv.2002.11054}
presents a new approach to building reusable and extensible compiler
infrastructure. MLIR is composed of a set of \emph{dialect} IRs, subsets
of which are mutually compatible, either directly or by way of translation/legalization.
The various dialects aim to capture and formalize the semantics of
compute intensive programs at varying levels of abstraction, as well
as namespace-related sets of IR transformations. The entrypoint into
this compiler framework from PyTorch is the \texttt{torch} dialect~\cite{torch-mlir}, a high-fidelity mapping from TS IR to MLIR native IR, which, in addition to performing the translation to MLIR, fully
refines all shapes of intermediate tensors in the DNN (i.e., computes
concrete values for all dimensions of each tensor), a necessary step
for downstream optimizations and eliminating inconsistencies in the
DNN~\cite{https://doi.org/10.48550/arxiv.2203.08402}.

While necessary for lowering to MLIR
and shape refinement,  the \texttt{torch} dialect represents a DNN at the same
level of abstraction as TS IR: it does not capture the precise data and control flow needed for de novo implementations of DNN
operations (e.g., for FPGA). Fortunately, MLIR supports lower-level
dialects, such as \texttt{linalg}, \texttt{affine}, and \texttt{scf}. The \texttt{scf} (structured control flow) dialect describes
standard control flow primitives, such as conditionals and loops,
and is mutually compatible with the \texttt{arith} (arithmetic operations)
and \texttt{memref} (memory buffers) dialects. The \texttt{affine}
dialect, on the other hand, provides a formalization of semantics
that lend themselves to polyhedral compilation techniques~\cite{polyhedral-mlir} that enable loop dependence analysis and loop transformations.
Such loop transformations, particularly loop unrolling, are crucial
for achieving lowest possible latencies~\cite{yehpca2022scalehls}
because loop nests directly inform the concurrency and parallelism
of the final RTL design.

\subsection{High-level synthesis}
\label{subsec:High-level-synthesis}

High-level synthesis tools produce RTL descriptions of designs from
high-level representations, such as C or C++~\cite{10.1145/2514740,ferrandi2021bambu}.
In particular, Xilinx's Vitis HLS, based on the Autopilot project~\cite{Zhang2008}, is a state-of-the-art HLS tool. Given a high-level,
procedural, representation, HLS carries out three fundamental tasks,
in order to produce a corresponding RTL design:
\begin{enumerate}
\item HLS schedules operations (such as \texttt{mulf}, \texttt{addf}, \texttt{load},
\texttt{store}) in order to determine which operations should occur
during each clock cycle; such a schedule depends on three characteristics
of the high-level representation:
(a) the topological ordering of the DFG of the procedural representation
(i.e., the dependencies of operations on results of other operations
and resources);
(b) the delay for each operation; and
(c) the user's desired clock rate/frequency.
\item HLS associates (\emph{binds}) floating point operations to
RTL instantiations of intellectual property (IP) for those operations;
for example whether to associate an addition operation followed by
a multiply operation to IPs for each, or whether to associate them
both with a single IP, designed to perform a fused multiply-accumulate
(MAC). In the case of floating-point arithmetic operations, HLS also (with
user guidance) determines the precision of the floating-point representation.
\item HLS builds a finite-state machine (FSM) that implements the schedule
of operations as control logic, i.e., logic that initiates operations
during the appropriate stages of the schedule.
\end{enumerate}
In addition to fulfilling these three fundamental tasks, HLS aims to optimize the program. In particular, HLS attempts
to maximize concurrency and parallelism (number of concurrent operations
scheduled during a clock cycle) in order maximize the throughput and
minimize the latency of the final implementation. Maximizing concurrency
entails pipelining operations: operations are executed such that they
overlap in time when possible, subject to available resources. Maximizing
parallelism entails partitioning the DNN into subsets of operation
that can be computed independently and simultaneously and whose results
are aggregated upon completion.

\begin{listing}
\begin{minted}[escapeinside={||},mathescape=true]{python}
def conv2d(
  input: MemRef(|$b$|, |$c_{in}$|, |$h$|, |$w$|),
  output: MemRef(|$b$|, |$c_{out}$|, |$h$|, |$w$|),
  weight: MemRef(|$c_{out}$|, |$c_{in}$|, |$k$|, |$k$|)
):
  for i1 in range(0, |$b$|):
    for i2 in range(0, |$c_{out}$|):
      for i3 in range(0, |$h$|):
        for i4 in range(0, |$w$|):
          for i5 in range(0, |$c_{in}$|):
            for i6 in range(0, |$k$|):
              for i7 in range(0, |$k$|):
                _3 = i3 + i6
                _4 = i4 + i7
                _5 = input[i1, i5, _3, _4]
                _6 = weight[i2, i5, i6, i7]
                _7 = output[i1, i2, i3, i4]
                _8 = _5 * _6
                _9 = _7 + _8
                output[i1, i2, i3, i4] = _9
\end{minted}
\caption{Python representation of a padding $\left\lfloor k/2\right\rfloor $,
stride 1, $c_{out}$ filter convolution with $k\times k$ kernel applied
to ($\ensuremath{b},\ensuremath{c_{in}},\ensuremath{h},\ensuremath{w}$)-dimensional\texttt{
input} tensor; $b$ is batch size, $c_{in}$ is number
of channels, and ($h,w$) are height and width, respectively.\label{lis:Single-filter-convolution}}
\end{listing}

While HLS aims to optimize various characteristics of
a design automatically, there are challenges associated this automation. In particular, maximum concurrency and parallelism necessitates
data-flow analysis in order to identify data dependencies amongst
operations, both for scheduling and identifying potential data hazards.
Such data-flow analysis is expensive and grows (in runtime) as better
performance is pursued. This can be understood in terms of loop-nest
representations of DNN operations.

For example, consider the convolution
in Listing~\ref{lis:Single-filter-convolution}.
A schedule that parallelizes (some of) the arithmetic operations
for this loop nest can be computed by first unrolling the loops up
to some ``trip count'' and then computing the topological sort of
the operations. When using this \emph{list
scheduling} algorithm, the degree to which the loops are unrolled determines
how many arithmetic operations can be scheduled in parallel. The issue
is that the \texttt{store}s and \texttt{load}s on the \texttt{output}
array prevent reconstruction of explicit relationships between the
inputs and outputs of the arithmetic operations across loop iterations.
The conventional resolution to this loss of information is to perform
\emph{store-load forwarding}: pairs of \texttt{store} and \texttt{load}
operations on the same memory address are eliminated, with the operand
of the \texttt{store} forwarded to the uses of the \texttt{load} (see
Listing~\ref{lis:Single-filter-convolution-1}).
\begin{listing}
\begin{minted}[numbers=left,escapeinside={||},mathescape=true,highlightlines={19,25},autogobble=true,numbersep=3pt]{python}
def conv2d(
  input: MemRef(|$b$|, |$c_{in}$|, |$h$|, |$w$|),
  output: MemRef(|$b$|, |$c_{out}$|, |$h$|, |$w$|),
  weight: MemRef(|$c_{out}$|, |$c_{in}$|, |$k$|, |$k$|)
):
  for i1 in range(0, |$b$|):
    for i2 in range(0, |$c_{out}$|):
      for i3 in range(0, |$h$|):
        for i4 in range(0, |$w$|):
	  ...
	  # e.g., i5, i6, i7 = 2, 3, ${\setlength{\fboxsep}{1pt}\colorbox{Salmon}{\texttt{4}}}$
	  _31 = i3 + i6
	  _41 = i4 + i7
	  _51 = input[i1, i5, _31, _41]
	  _61 = weight[i2, i5, i6, i7]
	  _71 = output[i1, i2, i3, i4]
	  _81 = _51 * _61
	  |${\setlength{\fboxsep}{1pt} \colorbox{green}{\texttt{\_91}}}$| = _71 + _81
          |${\setlength{\fboxsep}{1pt}          \colorbox{green}{\texttt{output[i1, i2, i3, i4]}}}$| = |${\setlength{ \fboxsep}{1pt} \colorbox{green}{\texttt{\_91}}}$|
	  # i5, i6, i7 = 2, 3, ${\setlength{\fboxsep}{1pt}\colorbox{Salmon}{\texttt{5}}}$
	  _32 = i3 + i6
	  _42 = i4 + i7
	  _52 = input[i1, i5, _32, _42]
	  _62 = weight[i2, i5, i6, i7]
	  |${\setlength{\fboxsep}{1pt}\colorbox{yellow}{\texttt{\_72}}}$| = |${\setlength{\fboxsep}{1pt}          \colorbox{green}{\texttt{output[i1, i2, i3, i4]}}}$|
	  _82 = _52 * _62
	  |${\setlength{\fboxsep}{1pt}\colorbox{Cyan}{\texttt{\_92}}}$| = |${\setlength{\fboxsep}{1pt}\colorbox{yellow}{\texttt{\_72}}}$| + _82
	  output[i1, i2, i3, i4] = _92
	  ...
\end{minted}
\caption{Store-load forwarding across successive iterations (e.g.,\texttt{
i7} $={\setlength{\fboxsep}{1pt}\colorbox{Salmon}{\texttt{4}}}, {\setlength{\fboxsep}{1pt}\colorbox{Salmon}{\texttt{5}}}$)
of the inner loop in Listing~\ref{lis:Single-filter-convolution},
after unrolling. The forwarding opportunity is from the store on line
19 to the load on line 25; both can be eliminated and ${\setlength{\fboxsep}{1pt}\colorbox{green}{\texttt{\_91}}}$
can replace uses of ${\setlength{\fboxsep}{1pt}\colorbox{yellow}{\texttt{\_72}}}$,
such as in the computation of \texttt{${\setlength{\fboxsep}{1pt}\colorbox{Cyan}{\texttt{\_92}}}$}
(and potentially many others).\label{lis:Single-filter-convolution-1}}
\end{listing}
Ensuring correctness of this transformation (i.e., that it preserves program
semantics) requires verifying, for each pair of candidate \texttt{store} and \texttt{load}
operations, that there is no intervening memory
operation on the same memory address. These verifications are non-trivial
since the iteration spaces of the loops need not be regular; in general
it might involve solving a small constraint satisfaction program~\cite{rajopadhye2002dependence}.
Furthermore, the number of required verifications grows polynomially in
the convolution parameters, since the loop nest unrolls into
$b\times c_{out}\times h\times w\times c_{in}\times k^{2}$ \texttt{store}-\texttt{load}
pairs on the \texttt{output} array.

Finally, note, although greedy solutions to the scheduling problem solved
by HLS are possible, the scheduling problem, in principle, can be formulated
as an integer linear program (ILP), for which the corresponding decision problem is complete for NP.
In summary, HLS tools solve computationally intensive problems in
order to produce an RTL description of a high-level representation
of a DNN. These phases of the HLS process incur ``development time''
costs (i.e., runtime of the tools) and impose practical limitations
on the amount of design space exploration (for the purpose of achieving
latency goals) which can be performed. \texttt{OpenHLS} addresses
these issues by enabling the user to employ heuristics during both
the parallelization and scheduling phases which, while not guaranteed
to be correct (but can be \emph{behaviorally verified}) and have much
lower runtimes (see Section~\ref{subsec:Symbolic-execution-for}).

\subsection{FPGA design}

Broadly, at the register-transfer level of abstraction, there remain
two more steps prior to being able to deploy a design to
an FPGA: a final lowering, so-called logic synthesis,
and place and route (P\&R). The entire process may be
carried out by Xilinx's Vivado tool.

Logic synthesis is the process of mapping RTL to actual hardware primitives
on the FPGA (so-called \emph{technology mapping}), such as lookup
tables (LUTs), block RAMs (BRAMs), flip-flops (FFs), and digital signal
processors (DSPs). Logic synthesis produces a network list (\emph{netlist})
describing the logical connectivity of various parts of the design.
Logic synthesis, for example, determines the implementation of floating-point
operations in terms of DSPs; depending on user parameters and other
design features, DSP resource consumption for floating-point multiplication
and addition can differ greatly. Logic synthesis also determines the
number of LUTs and DSPs which a high-level representation of a DNN
corresponds to, which is relevant to both the performance and feasibility of that DNN when deployed to FPGA.

After the netlist has been produced, the entire design undergoes P\&R to determine which configurable logic block within
an FPGA should implement each of the units of logic required by the
digital design. P\&R algorithms need to minimize distances between
related units of functionality (in order to minimize wire delay),
balance wire density across the entire fabric of the FPGA (in order
to reduce route congestion), and maximize the clock speed of the design
(a function of both wire delay, logic complexity, and route congestion).
The final, routed design, can then be deployed to the FPGA by producing
a proprietary \emph{bitstream}, which configures the FPGA.

\subsection{Related work\label{subsec:relatedwork}}

Several projects aim to support translation from high-level representations of DNNs to feasible FPGA designs. Typically they rely on commercial HLS tools for the scheduling, binding, and RTL emission phases of the translation, such as in the cases of DaCeML~\cite{daceml}, hls4ml~\cite{Duarte_2018}, and ScaleHLS~\cite{yehpca2022scalehls}, which all rely on Xilinx's Vitis HLS. Thus, they fail to efficiently (i.e., without incurring the aforementioned runtime costs) produce feasible and low-latency designs. One notable recent work is the SODA Synthesizer~\cite{9786533}, which does not rely on a commercial tool but instead relies on the open-source PandA-Bambu HLS tool~\cite{ferrandi2021bambu}; though open-source and mature, we found in our own tests that PandA-Bambu also could not handle fully unrolled designs efficiently.

Alternatively, some projects do not rely on HLS for scheduling, binding, and RTL emission, and also attempt to translate from high-level representations of DNNs to feasible FPGA designs, such as DNN Weaver~\cite{7783720} and NNGen~\cite{takamaeda2015pyverilog}. Both of the cited projects function as parameterized/templatized RTL generators and thus lack sufficient generality for our needs; primarily they seek to produce implementations of kernels that emulate GPU architectures (i.e., optimizing for throughput rather than latency).
In our experiments they were unable to generate low-latency implementations, either by achieving unacceptable latencies or by simply failing outright. (NNGen, due to the nature of templates, supports only limited composition, and produced ``recursion'' errors.)

\section{The Compiler and HLS framework\label{sec:OpenHLS-compiler-and}}

\begin{figure}[tbh]
\centering{}\includegraphics[width=1\columnwidth]{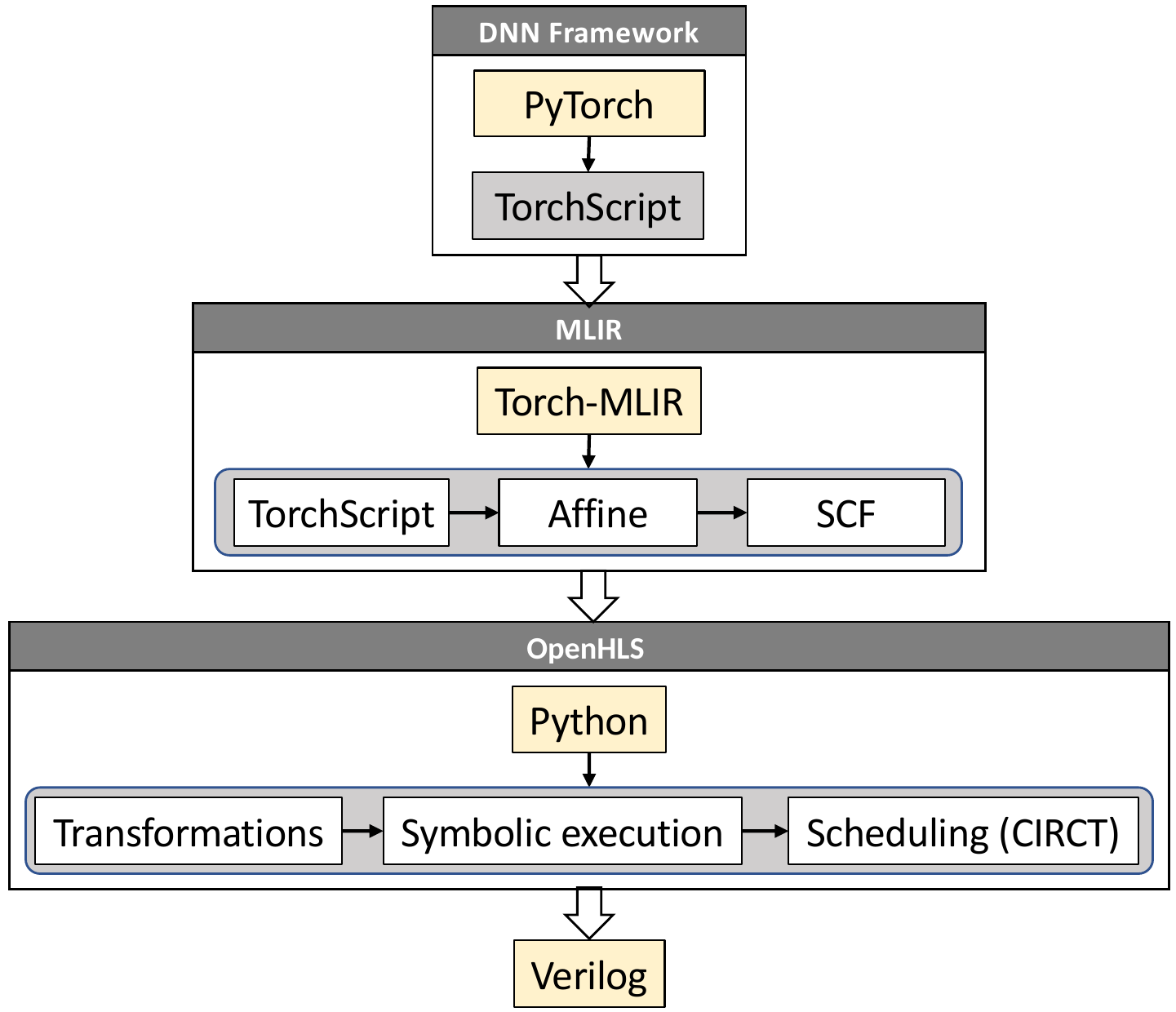}\caption{\texttt{OpenHLS} framework overview.\label{fig:OpenHLS-framework-overview.-3}}
\end{figure}

\texttt{OpenHLS} is an open source compiler and HLS framework that employs MLIR
for extracting loop-nest representations of DNNs. Implemented
in Python for ease of use and extensibility, it handles the DNN transformations as well as scheduling, binding, and FSM extraction. Importantly, there is no dependence on commercial HLS tools, a property that uniquely enables its use for applications that require the flexibility of open source tool (e.g., the ability to inspect and modify internals in order to adapt to special cases), such as low-latency physical science experiments. 
Figure~\ref{fig:OpenHLS-framework-overview.-3} shows its overall architecture.
\texttt{OpenHLS} first lowers DNNs from PyTorch to
MLIR through TorchScript and the \texttt{torch} dialect (see Section
\ref{subsec:MLIR}) and then from the \texttt{torch}
dialect to the \texttt{scf} dialect (through the \texttt{linalg} dialect).
Such a representation lends itself to a straightforward translation
to Python (compare Listing~\ref{lis:Single-filter-convolution} to
Listing~\ref{lis:Single-filter-convolution-2-1}) and indeed \texttt{OpenHLS}
performs this translation.


\begin{listing}
\begin{minted}[escapeinside={||},mathescape=true]{mlir}
@conv2d(
    %input: memref<|$b \times c_{in} \times h \times w$|>,
    %weight: memref<|$b \times c_{out} \times h \times w$|>,
    %output: memref<|$c_{out} \times c_{in} \times k \times k$|>
) {
  scf.for %i1 = %c0 to |$b$| step %c1 {
    scf.for %i2 = %c0 to |$c_{out}$| step %c1 {
      scf.for %i3 = %c0 to |$h$| step %c1 {
        scf.for %i4 = %c0 to |$w$| step %c1 {
          scf.for %i5 = %c0 to |$c_{in}$| step %c1 {
            scf.for %i6 = %c0 to |$k$| step %c1 {
              scf.for %i7 = %c0 to |$k$| step %c1 {
                %3 = arith.addi %i3, %i6
                %4 = arith.addi %i4, %i7
                %5 = memref.load %input[
                  %i1, %i5, %i3, %3, %4]
                %6 = memref.load %weight[
                  %i2, %i5, %i6, %i7]
                %7 = memref.load %output[
                  %i1, %i2, %i3, %i4]
                %8 = arith.mulf %5, %6
                %9 = arith.addf %7, %8
                memref.store %9, %output[
                  %i1, %i2, %i3, %i4]
              }
            }
          }
        }
      }
    }
  }
  return %2
}
\end{minted}
\caption{\texttt{scf} dialect loop representation of 
Listing
\ref{lis:Single-filter-convolution}.\label{lis:Single-filter-convolution-2-1}}
\end{listing}
The benefits of translating \texttt{scf} dialect to Python are manifold: see Section~\ref{subsec:Symbolic-execution-for}.
Ultimately, \texttt{OpenHLS} produces a representation of the DNN
that is then fully scheduled by using the scheduling infrastructure in
CIRCT~\cite{oppermann2022eurollvm} (an MLIR adjacent project). After
scheduling, \texttt{OpenHLS} emits corresponding RTL (as Verilog).

\texttt{OpenHLS} delegates to the FloPoCo~\cite{8877424} IP generator
the task of generating pipelined implementations of the standard floating-point
arithmetic operations (\texttt{mulf}, \texttt{divf}, \texttt{addf},
\texttt{subf}, \texttt{sqrtf}) at various precisions. In addition,
we implement a few generic (parameterized by bit width) operators
in order to support a broad range of DNN operations: two-operand maximum
(\texttt{max}), unary negation (\texttt{neg}), and the rectified linear
unit (\texttt{relu}). Transcendental functions, such as \texttt{exp},
are implemented by using a Taylor series expansion to $k$-th order (where
$k$ is determined on a case-by-case basis). Note that FloPoCo's floating-point
representation differs slightly from IEEE754, foregoing subnormals
and differently encoding zeroes, infinities and NaNs (for the benefit
of reduced complexity) and our implementations \texttt{max}, \texttt{neg},
\texttt{relu} are adjusted appropriately.

We now discuss some aspects of \texttt{OpenHLS} in more detail.

\subsection{Symbolic interpretation for fun and profit\label{subsec:Symbolic-execution-for}}

\begin{figure}[tbh]
\begin{centering}
\includegraphics[width=1\columnwidth,trim=0 7mm 0 6.5mm,clip]{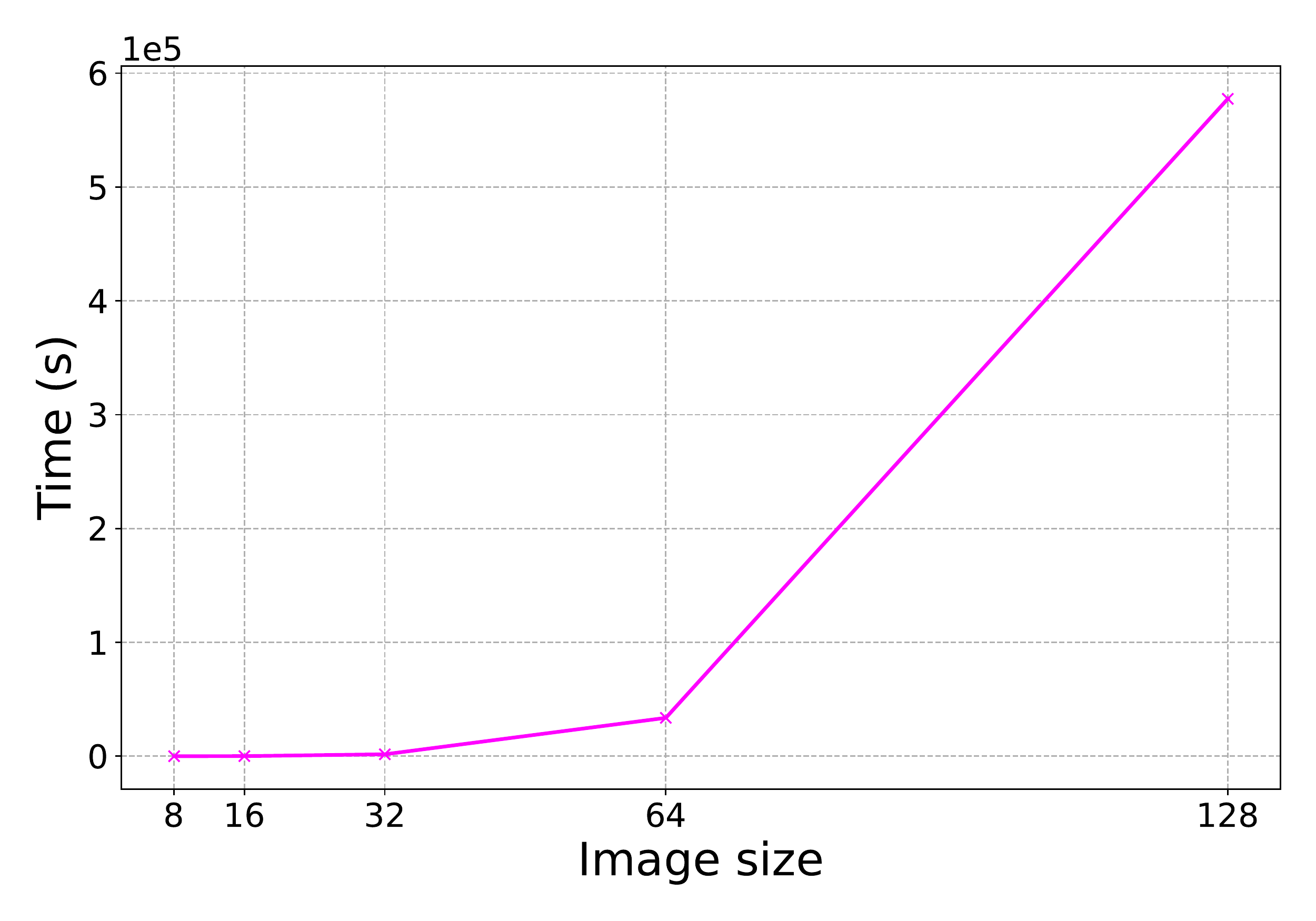}
\par\end{centering}
\caption{3$\times$3-kernel convolution (cf. Listing~\ref{lis:Single-filter-convolution-2-1})
full unrolling time vs. input (square) image size, with \texttt{store}-\texttt{load}
forwarding using MLIR's \texttt{-affine-scalrep} pass. The
longest time is 577,419 s ($\approx$160~h) for a loop nest
with a trip count of 128$\times$128$\times$3$\times$3=147,456. 
\label{fig:-kernel-convolution-full}
}
\end{figure}

As noted in Section~\ref{subsec:High-level-synthesis}, maximizing
concurrency and parallelism for a design entails unrolling loops
and analyzing the data flow of their operations. As illustrated in Figure~\ref{fig:-kernel-convolution-full}, the formally correct approach to unrolling loop nests can be prohibitively expensive in terms of runtime. In the
case of \texttt{BraggNN} (see Listing~\ref{lis:braggnn}), for example, the high cost of unrolling precluded effective search of the design space for a RTL representation
achieving the target latency.
Translating \texttt{scf} dialect to Python enables \texttt{OpenHLS}
to overcome this barrier by enabling us to use the Python interpreter
as a \emph{symbolic interpreter}. Interpreting the resulting Python
loop nests (i.e., running the Python program) while treating the arithmetic
and memory operations on SSA values as operations on symbols (i.e.,
Python classes with overloaded methods) enables us to:
\begin{enumerate}
\item Partially evaluate functions of iteration variables (for example,  \mintinline{mlir}!
to determine array index operands of all stores
and loads (for example, \linebreak \mintinline[fontsize={\small}]{mlir}!memref.load 
and thereupon \linebreak perform memory dependence checks, thus transforming
the problem of statically verifying memory dependence into one
of checking assertions at runtime;
\item Unroll loops by recording each floating-point arithmetic operation
executed while enforcing SSA; e.g., for a loop whose body has repeated
assignments to the same SSA value (ostensibly violating SSA), we execute
the loop and instantiate new, uniquely identified, symbols for the
result of each operation;
\item Reconstruct all data flow through arithmetic operations and memory
operations by interpreting \texttt{memref}s as \emph{geometric symbol
tables} (i.e., symbol tables indexed by array indices rather than
identifiers/names) and \texttt{store}s and \texttt{load}s as reads
and writes on those symbol tables;
\item Swap evaluation rules in order to support various functional
modes, e.g., evaluating floating-point arithmetic operations by using
(Python) bindings to FloPoCo's C++ functional models, thereby enabling
behavioral verification of our designs.
\end{enumerate}

See Table~\ref{tab:scf-dialect-to} for the translation rules from MLIR dialects to Python.

{\scriptsize{}}
\begin{figure*}[tbh]
\begin{centering}
\begin{tabular}{ll}
\toprule
$\left\llbracket \texttt{MLIR}\right\rrbracket $ & {\small{}Python}\tabularnewline
\midrule
\mintinline{mlir}!|$\llbracket$|
\addlinespace[1mm]
\mintinline{mlir}!|$\llbracket$|memref<|$b\times c_{in}\times h\times w$|>|$\rrbracket$|! & \mintinline{python}!MemRef(|$b$|, |$c_{in}$|, |$h$|, |$w$|)!\tabularnewline
\addlinespace[1mm]
\mintinline{mlir}!|$\llbracket$||$\perc$|5 = memref.load 
\addlinespace[1mm]
\mintinline{mlir}!|$\llbracket$| memref.store 
\addlinespace[1mm]
\mintinline{mlir}!|$\llbracket$| scf.for 
\addlinespace[1mm]
\mintinline{mlir}!|$\llbracket$||$\perc$|3 = arith.addi 
\addlinespace[1mm]
\mintinline{mlir}!|$\llbracket$||$\perc$|8 = arith.mulf 
\addlinespace[1mm]
\mintinline{mlir}!|$\llbracket$||$\perc$|9 = arith.addf 
\addlinespace[1mm]
\multicolumn{2}{c}{%
\begin{tabular}{ccc}
\multicolumn{1}{c}{\mintinline{mlir}!|$\llbracket$||$\perc$|63 = arith.cmpfugt |$\perc$|10, |$\perc$|c0|$\rrbracket$|!} & $\wedge$ & \multicolumn{1}{c}{\mintinline{mlir}!|$\llbracket$||$\perc$|64 = arith.select |$\perc$|63, |$\perc$|10, |$\perc$|c0|$\rrbracket$|!}\tabularnewline
\noalign{\vskip1mm}
\hline
\noalign{\vskip1mm}
\multicolumn{3}{c}{\mintinline{python}!|$\llbracket$||$\perc$|64|$\rrbracket$|.relu(|$\llbracket$||$\perc$|10|$\rrbracket$|)!}\tabularnewline
\end{tabular}}\tabularnewline
\addlinespace[1mm]
\multicolumn{2}{c}{%
\begin{tabular}{ccc}
\multicolumn{1}{c}{\mintinline{mlir}!|$\llbracket$||$\perc$|8 = arith.mulf |$\perc$|5, |$\perc$|6|$\rrbracket$|!} & $\wedge$ & \multicolumn{1}{c}{\mintinline{mlir}!|$\llbracket$||$\perc$|9 = arith.addf |$\perc$|7, |$\perc$|8|$\rrbracket$|!}\tabularnewline
\noalign{\vskip1mm}
\hline
\noalign{\vskip1mm}
\multicolumn{3}{c}{\mintinline{python}!|$\llbracket$||$\perc$|9|$\rrbracket$| = fma(|$\llbracket$||$\perc$|5|$\rrbracket$|, |$\llbracket$||$\perc$|6|$\rrbracket$|, |$\llbracket$||$\perc$|7|$\rrbracket$|)!}\tabularnewline
\end{tabular}}\tabularnewline
\bottomrule
\addlinespace[1mm]
\end{tabular}{\scriptsize\par}
\par\end{centering}
{\scriptsize{}\caption{Translation rules for mapping \texttt{scf}, \texttt{arith}, and \texttt{memref} dialects to Python.\label{tab:scf-dialect-to}}
}%

\end{figure*}
{\scriptsize\par}

\subsection{AST transformations and 
verification\label{subsec:AST-transformations-and-1}}

Prior to interpretation, \texttt{OpenHLS} performs some simple AST
transformations on the Python generated from \texttt{scf} dialect:
\begin{enumerate}

\item \textbf{Hoist globals}: Move fixed DNN tensors (i.e., weights) out of the body of the generated Python function (\texttt{OpenHLS} translates the MLIR \texttt{module} corresponding
to the DNN into a single Python function in order to simplify analysis
and interpretation) and into the parameter list, for the purpose of ultimately exposing
them at the RTL module interface.

\item \textbf{Remove }\texttt{\textbf{if}}\textbf{ expressions}: DNN \texttt{relu}
operations are lowered to the \texttt{scf} dialect as a decomposition
into \texttt{arith.cmpfugt} and \texttt{arith.select}; this transformation
recomposes them into a \texttt{relu}.

\item \textbf{Remove MACs}: Schedule sequences of \texttt{load}-\texttt{multiply}-\texttt{add}-\texttt{store}
(common in DNN implementations) 
jointly, coalescing them into a single
\texttt{fmac} operation.

\item \textbf{Reduce }\texttt{\textbf{for}}\textbf{s}:
Implement the reduction tree structure for non-parallelizable loop nests
mentioned in Section~\ref{subsec:Scheduling}.
\end{enumerate}

These transformations on the Python AST are simple (implemented with
procedural pattern matching), extensible, and efficient (marginal
runtime cost) because 
no effort is made to verify
their formal correctness. Thus, \texttt{OpenHLS} trades formal correctness
for development time performance. This tradeoff enables quick design
space iteration, which for example, enabled us to achieve low
latency implementations for \texttt{BraggNN} (see Section~\ref{sec:BraggNN-case-study}).

\texttt{OpenHLS} supports
behavioral rather than formal verification. Specifically, \texttt{OpenHLS} can generate
testbenches for all synthesized RTL. The test vectors for these testbenches
are generated by evaluating the generated Python representation of
the DNN on randomly generated inputs but with floating-point operations
now evaluated using functional models of the corresponding FloPoCo
operators. The testbenches can then be run using any IEEE 1364 compliant
simulator. We run a battery of such testbenches (corresponding
to various DNN operation types), using \texttt{cocotb}~\cite{rosser2018cocotb}
and \texttt{iverilog}~\cite{williamsicarus}, as a part of our continuous integration (CI) process.

\subsection{Scheduling\label{subsec:Scheduling}}

Recall that HLS must schedule operations during each clock cycle in a way that
preserves the DNN's data-flow graph. 
That schedule then informs
the construction of a corresponding FSM. As already mentioned, scheduling
an arbitrary DNN involves formulating and solving an ILP. 
In the resource-unconstrained
case, due to the precedence relations induced by data flow, the constraint
matrix of the associated ILP is a \emph{totally unimodular matrix}
and the feasible region of the problem is an integral polyhedron.
In such cases, the scheduling problem can be solved optimally
in polynomial time with a LP solver~\cite{tuprints9272}.
In the resource-constrained
case, resource constraints can also be transformed into precedence
constraints by picking a particular (possibly heuristic)
linear ordering on the resource-constrained operations. This transformation
partitions resource-constrained operations into distinct clock cycles,
thereby guaranteeing sufficient resources are available for all operations
scheduled within the same clock cycle~\cite{10.1145/3174243.3174268}.

\texttt{OpenHLS} uses the explicit parallelism of the \texttt{scf.parallel}
loop-nest representation to inform such a linear ordering on resource-constrained
operations. By assumption, for loop nests which can be reprepresented
as \texttt{scf.parallel} loop nests (see Listing~\ref{lis:Single-filter-convolution-2-2}),
each instance of a floating-point arithmetic operation in the body
corresponding to unique values of the iteration variables (e.g., \texttt{\%i1},\texttt{
\%i2},\texttt{ \%i3},\texttt{ \%i4} for Listing~\ref{lis:Single-filter-convolution-2-2})
is independent of all other such instances, although data flow within a loop body must still be respected.
This exactly determines total resource usage per loop nest; for example,
the convolution in Listing~\ref{lis:Single-filter-convolution-2-2}
would bind to $2K_{i}$ DSPs (assuming \texttt{mulf}, \texttt{addf}
bind to one DSP each), where:
\begin{tabbing}
\ \ \ \ \ \ \ $K_{i}\   \coloneqq$ \= $\left|\left\{ \texttt{\%i1}=\texttt{\%c0}+\texttt{\%c1}\times\mathbb{N}\,\wedge\,\texttt{\%i1}<b\right\} \right|\ \times$\\
\>$\left|\left\{ \texttt{\%i2}=\texttt{\%c0}+\texttt{\%c1}\times\mathbb{N}\,\wedge\,\texttt{\%i2}<c_{out}\right\} \right|\ \times$\\
\>$\left|\left\{ \texttt{\%i3}=\texttt{\%c0}+\texttt{\%c1}\times\mathbb{N}\,\wedge\,\texttt{\%i3}<h\right\} \right|\ \times$\\
\>$\left|\left\{ \texttt{\%i4}=\texttt{\%c0}+\texttt{\%c1}\times\mathbb{N}\,\wedge\,\texttt{\%i4}<w\right\} \right|$
\end{tabbing}

\noindent
with $\texttt{\%c1}\times\mathbb{N}$ representing all multiples of
$\texttt{\%c1}$. That is to say, $K_{i}$ is the cardinality of the
cartesian product of the iteration spaces of the parallel iteration
variables.

\begin{listing}
\begin{minted}[fontsize={\small},escapeinside={||},mathescape=true]{mlir}
@conv2d(
    %input: memref<|$b \times c_{in} \times h \times w$|>,
    %weight: memref<|$b \times c_{out} \times h \times w$|>,
    %output: memref<|$c_{out} \times c_{in} \times k \times k$|>
) {
  scf.parallel (%i1, %i2, %i3, %i4) =
               (%c0, %c0, %c0, %c0) to
               (|$b$|, |$c_{out}$|, |$h$|, |$w$|) step
               (%c1, %c1, %c1, %c1) {
    scf.for %i5 = %c0 to |$c_{in}$| step %c1 {
      scf.for %i6 = %c0 to |$k$| step %c1 {
        scf.for %i7 = %c0 to |$k$| step %c1 {
          %3 = arith.addi %i3, %i6
          %4 = arith.addi %i4, %i7
          %5 = memref.load %input[%i1, %i5, %i3, %3, %4]
          %6 = memref.load %weight[%i2, %i5, %i6, %i7]
          %7 = memref.load %output[%i1, %i2, %i3, %i4]
          %8 = arith.mulf %5, %6
          %9 = arith.addf %7, %8
          memref.store %9, %output[%i1, %i2, %i3, %i4]
        }
      }
    }
  }
  return %2
}
\end{minted}

\vspace{-1ex}

\caption{Parallel loop representation of 
Listing~\ref{lis:Single-filter-convolution}\label{lis:Single-filter-convolution-2-2}, exhibiting explicitly 
the resource partitioning and ordering strategy we employ  to construct a feasible schedule of operations.}
\end{listing}

Defining $K\coloneqq\max_{i}K_{i}$ across all \texttt{scf.parallel}
loop nests, we can infer peak usage of any resource. Then, after indexing
available hardware resources $j=1,\dots,K$, we can bind the operations
of any particular loop nest. This leads to a linear ordering on resource-constrained
operations such that operations bound to the same hardware resource
index $j$ must be ordered according to their execution order during
symbolic interpretation.\footnote{\texttt{OpenHLS} only needs to construct a partial precedence ordering
$\texttt{op}_{a}<\texttt{op}_{b}$ for operations $\texttt{op}_{a},\texttt{op}_{b}$
which CIRCT then combines with the delays of the operations to construct
constraints such as $\texttt{start\_op}_{a}+\texttt{delay}_{a}\leq\texttt{start\_op}_{b}$.} Note that this ordering coincides with the higher-level structure of
the DNN, which determines the ordering of\texttt{ scf.parallel} loop nests (and thus
interpretation order during execution of the Python program).

For DNN operations that lower to sequential loop nests rather than \texttt{scf.parallel} loop nests (e.g., \texttt{sum}, \texttt{max},
or \texttt{prod}), we fully unroll the loops and transform the resulting,
sequential, operations into a reduction tree; we use As-Late-As-Possible
scheduling~\cite{baruch1996scheduling} amongst the subtrees of such
reduction trees.

\section{Evaluation\label{sec:Evaluation}}

We evaluate \texttt{OpenHLS} both on individual DNN layers, and end-to-end,
on our use-case \texttt{BraggNN}. We compare \texttt{OpenHLS} to
Xilinx's Vitis HLS by comparing the latencies and resource usages
of the final designs generated by each. We also compare the runtimes
of the tools themselves. Both \texttt{OpenHLS} and Vitis HLS produce
Verilog RTL, on which we run a synthesis pass by using Xilinx's Vivado.
The particular FPGA target is Xilinx Alveo U280. We measure LUT, DSP,
BRAM, and FF usage. For the DNN layer evaluations, we use FloPoCo
(5,11)-floating point representations (5-bit
exponent, 11-bit mantissa), corresponding to Vitis HLS's
IEEE half-precision IPs. We synthesize all designs for a 10~ns target
clock period and report end-to-end latency as a product of the total
schedule interval count of the design and achieved clock period (\emph{10-WNS},
where \emph{WNS} is the worst negative slack reported). In the case
of Vitis HLS, which potentially explicitly pipelines the design and
therefore implements with an initiation interval strictly less than
the total schedule interval count, we report in terms of the best
possible interval count (\texttt{LatencyBest} from the Vitis HLS reports).
All other measurements are collected from Vivado synthesis reports.
As Vitis HLS operates on C++ representations, we generate
such a representation for our test cases by first lowering each DNN
layer to the \texttt{affine} dialect and then applying the \texttt{scalehls-translate}
tool of the ScaleHLS project~\cite{yehpca2022scalehls} to emit C++.
Importantly, we do not make any use of \texttt{scalehls-opt} optimization
tool (of the same project).

Since our ultimate goal is low latency inference, and since the strategy
that \texttt{OpenHLS} employs in the pursuit of this goal is loop
unrolling, in order to produce a like for like comparison, we similarly
unroll the representation that is passed to Vitis HLS. Thus, all Vitis
HLS measurements are reported in terms of \emph{unroll factor}: an
unroll factor of $k$ corresponds to a $k$-fold increase in the number
of statements in the body of a loop and commensurate $k$-fold decrease
in the trip count of the loop. For loop nests, we unroll inside out:
if $k$ is greater than the trip count $t$ of the innermost loop,
we unroll the innermost loop completely and then unroll the enclosing
loop by a factor of $k-t$. We do not perform any store-load forwarding
during this preprocessing but we annotate all arrays with the directive
\mintinline{tex}!array_partition complete dim=1! in order that Vitis
HLS can effectively pipeline. All representations generated by \texttt{OpenHLS}
correspond to full unrolling of the loop nests.

\subsection{DNN layers}

We evaluate \texttt{OpenHLS}
vs.\ Xilinx's Vitis HLS by comparing the latency of the final design
on five DNN layer types, chosen to cover a range of arithmetic operations (\texttt{mulf},
\texttt{divf}, \texttt{addf}, \texttt{subf}, \texttt{sqrtf}) and data
access patterns (iteration, accumulation, reduction):
\begin{itemize}
\item \mintinline{python}!addmm(a, b, c)!: Matrix multiply: 
\texttt{a} $\times$ \texttt{b} + \texttt{c};
\item \mintinline{python}!batch_norm_2d(num_features)!: Batch normalization
over a 4D input~\cite{https://doi.org/10.48550/arxiv.1502.03167};
\item \mintinline{python}!conv_2d(|$c_{in}$|, |$c_{out}$|, |$k$|)!: 2D
convolution with bias, with $k\times k$ kernel, over a $b\times c_{in}\times h\times w$
input, producing $b\times c_{out}\times h'\times w'$ output;
\item \mintinline{python}!max_pool_2d(|$k$|, stride)!: 2D max pooling,
with $k\times k$ kernel, and striding;
\item \mintinline{python}!soft_max!: 
$\begin{aligned}[t]
\text{softmax}\left(\boldsymbol{x}\right)\coloneqq\left[\frac{\exp\left(x_{i}\right)}{\sum_{j}\exp\left(x_{j}\right)}\right]
\end{aligned}$
\end{itemize}
The parameter values and input dimensions used during evaluation are
summarized in Table~\ref{tab:Resource-usage-for-2}.

\begin{table}[tbh]
\caption{DNN layers used for evaluation of \texttt{OpenHLS.} \label{tab:Resource-usage-for-2}}
\centering{}%
\begin{tabular}{lll}
\toprule
Layer & Parameter values & Input dimensions\tabularnewline
\midrule
\mintinline{python}!addmm! & N/A & 
$\texttt{a}, \texttt{b}, \texttt{c}: \left(16,16\right)$\tabularnewline
\midrule
\mintinline{python}!batch_norm_2d! & $\texttt{num\_features}=2$ & $\texttt{input}: \left(10,2,3,3\right)$\tabularnewline
\midrule
\mintinline{python}!conv_2d! & $c_{in}=1, c_{out}=k=3$ & $\texttt{input}: \left(1,1,16,16\right)$\tabularnewline
\midrule
\mintinline{python}!max_pool_2d! & $k=3, \texttt{stride}=2$ & $\texttt{input}: \left(1,3,16,16\right)$\tabularnewline
\midrule
\mintinline{python}!soft_max! & N/A & $\texttt{input}: \left(1,3,16,16\right)$\tabularnewline
\bottomrule
\end{tabular}
\end{table}

\begin{figure*}[!t]
\begin{centering}
\subfloat[\texttt{addmm} module]{\centering{}\includegraphics[width=1\columnwidth,trim=0 8mm 0 7mm,clip]{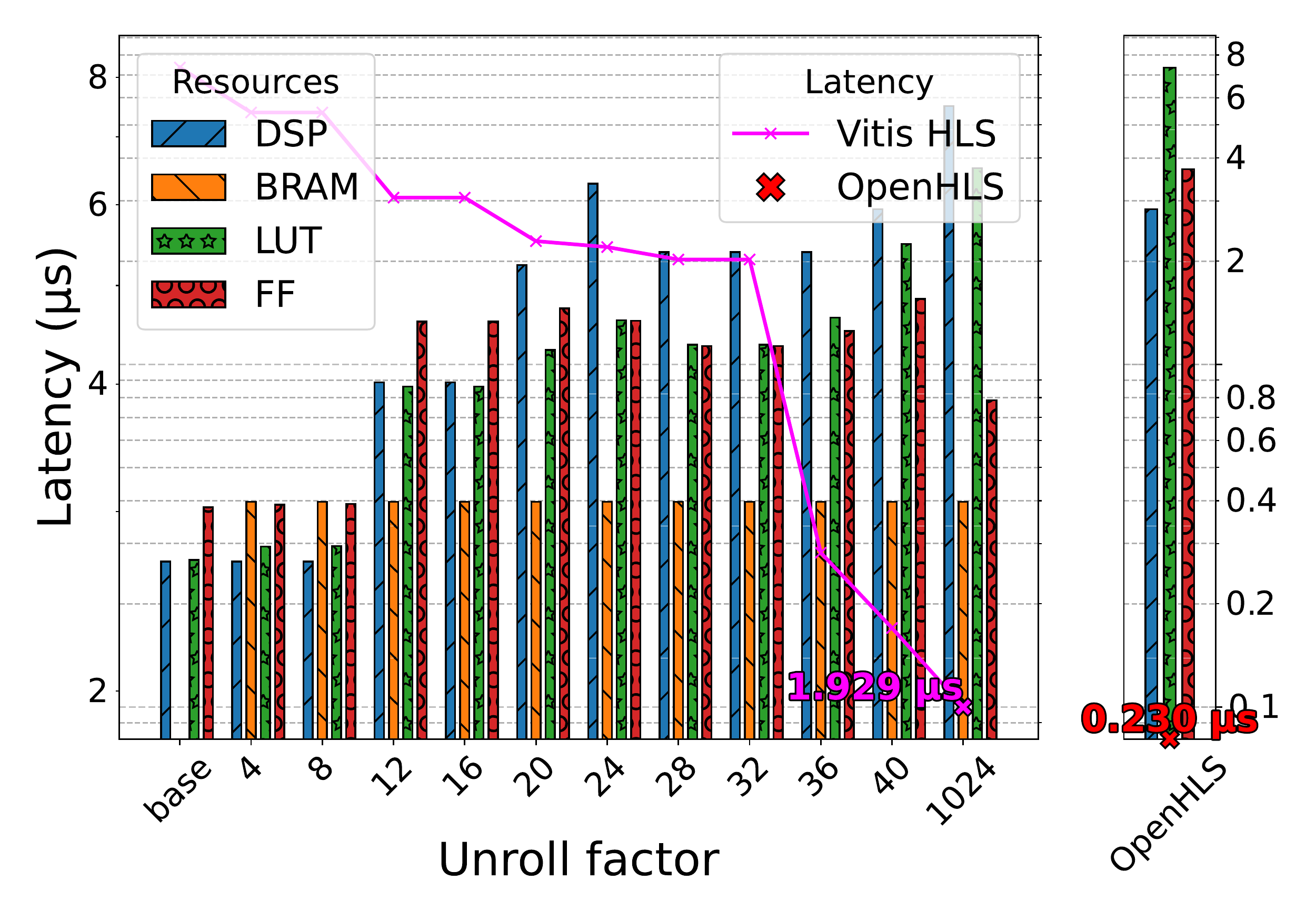}\label{2dlattice-1-1-2}}\subfloat[\texttt{batch\_norm\_2d} module]{\centering{}\includegraphics[width=1\columnwidth,trim=0 8mm 0 7mm,clip]{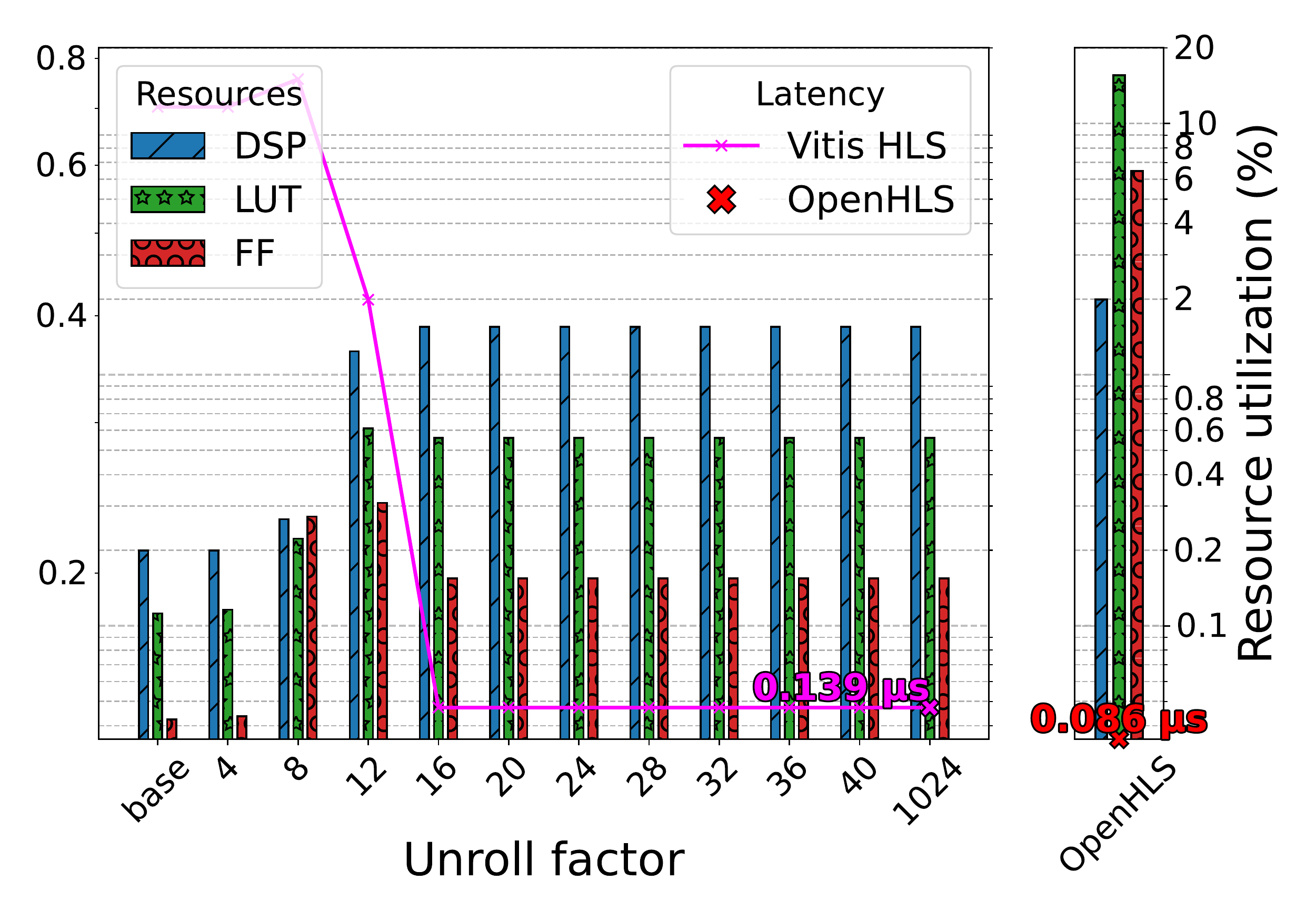}\label{2dlattice-1-2-2}}
\par\end{centering}
\medskip{}

\centering{}\label{2dlattice-1-3}\subfloat[\texttt{conv\_2d} module]{\centering{}\includegraphics[width=1\columnwidth,trim=0 8mm 0 7mm,clip]{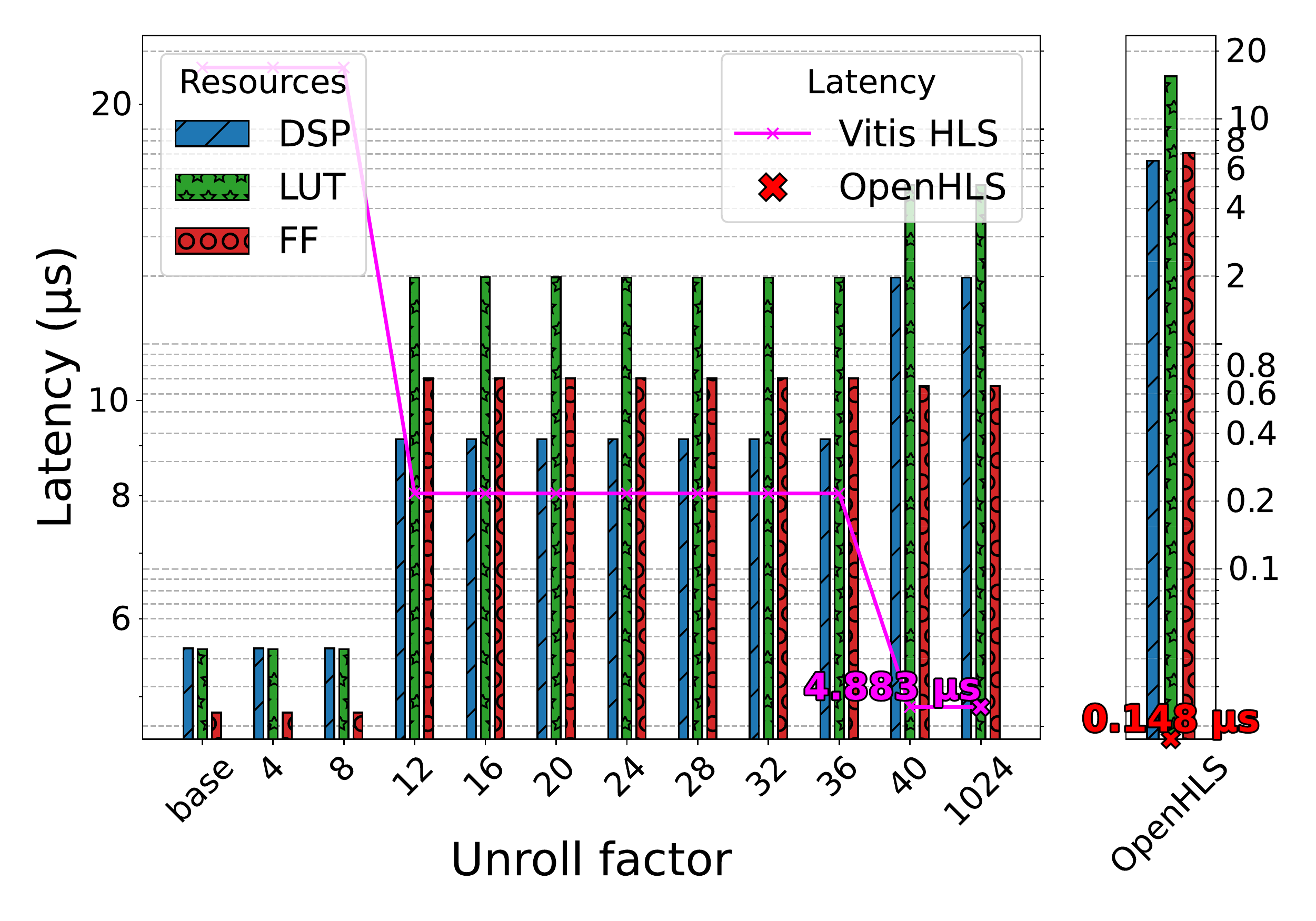}\label{2dlattice-1-1-1-1}}\subfloat[\texttt{max\_pool\_2d} module]{\centering{}\includegraphics[width=1\columnwidth,trim=0 8mm 0 7mm,clip]{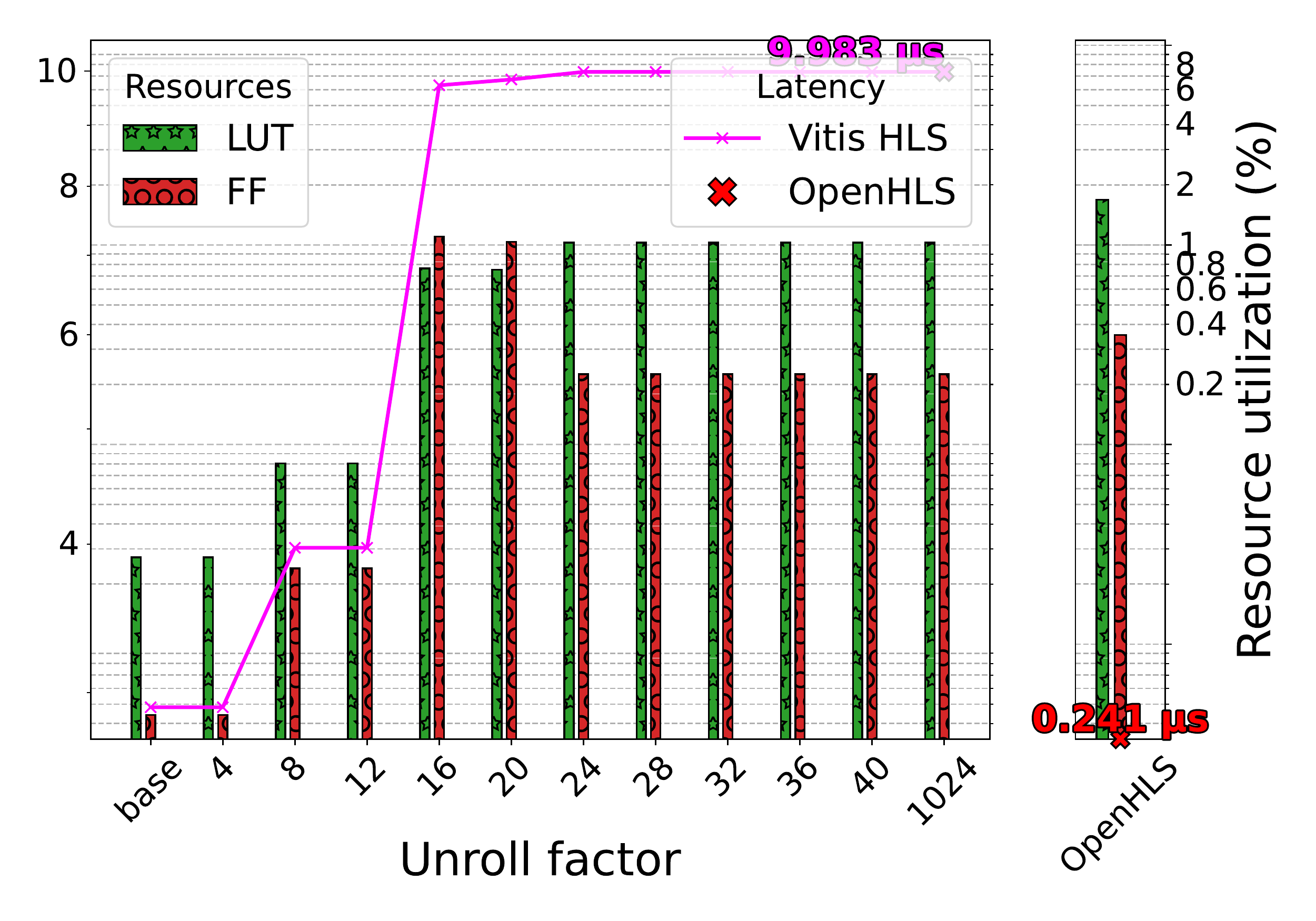}\label{2dlattice-1-2-1-1}}\medskip{}
\subfloat[\texttt{soft\_max} module]{\centering{}\includegraphics[width=1\columnwidth,trim=0 8mm 0 7mm,clip]{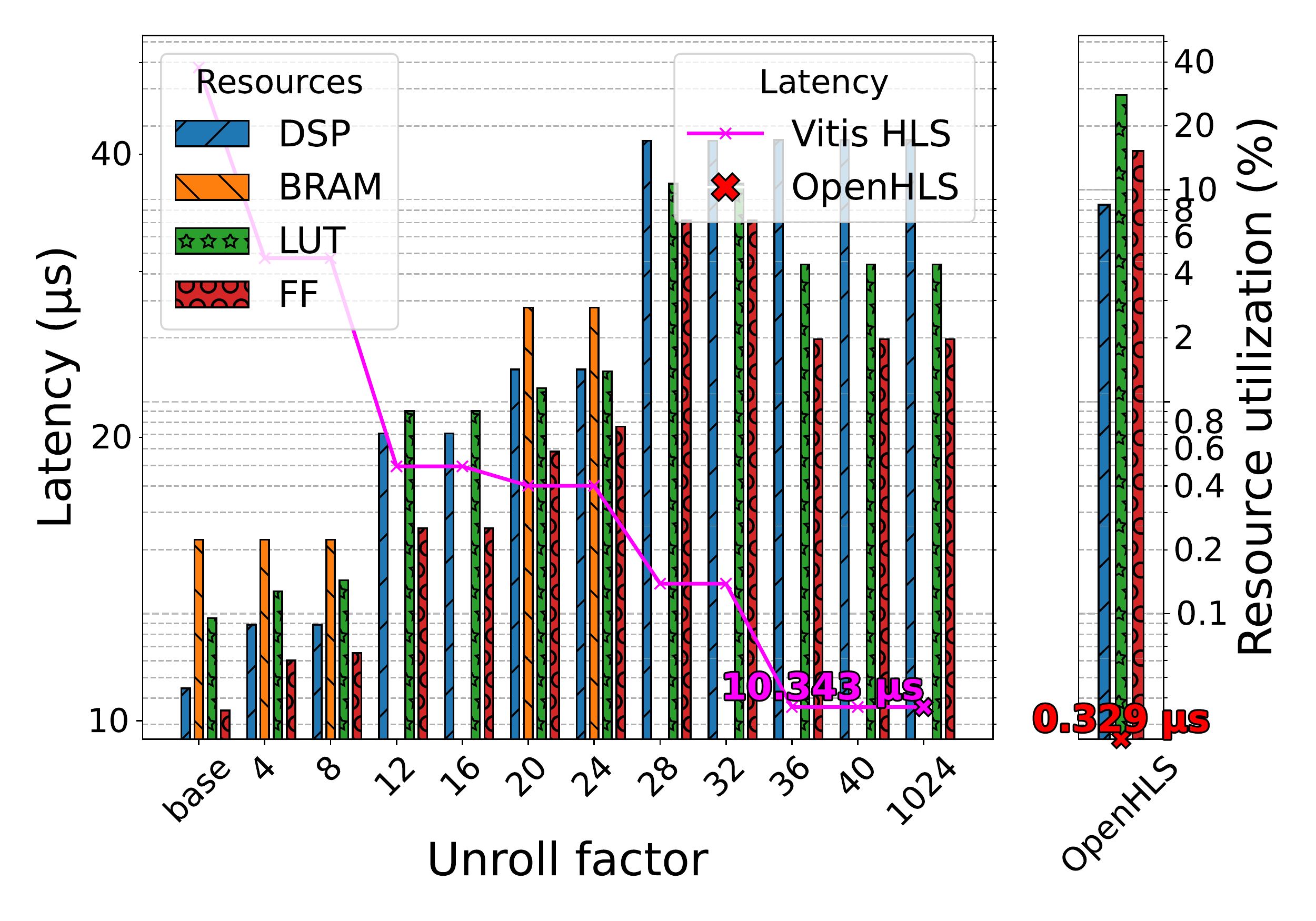}\label{2dlattice-1-2-1-1-1}}\caption{Vitis HLS vs. \texttt{OpenHLS} resource usage and latency vs. unroll
factor for five DNN modules, exhibiting the large runtime cost incurred in using Vitis HLS to search the design space (of possible low-latency designs for each layer). 
The lines give latencies (left axes); the bars give the \% of the resource used (right axes).
All \emph{y}-scales are log.
\label{fig:Resource-usage-and}}
\end{figure*}

\begin{figure}[tbh]
\centering{}\includegraphics[width=1\columnwidth,trim=0 8mm 0 7mm,clip]{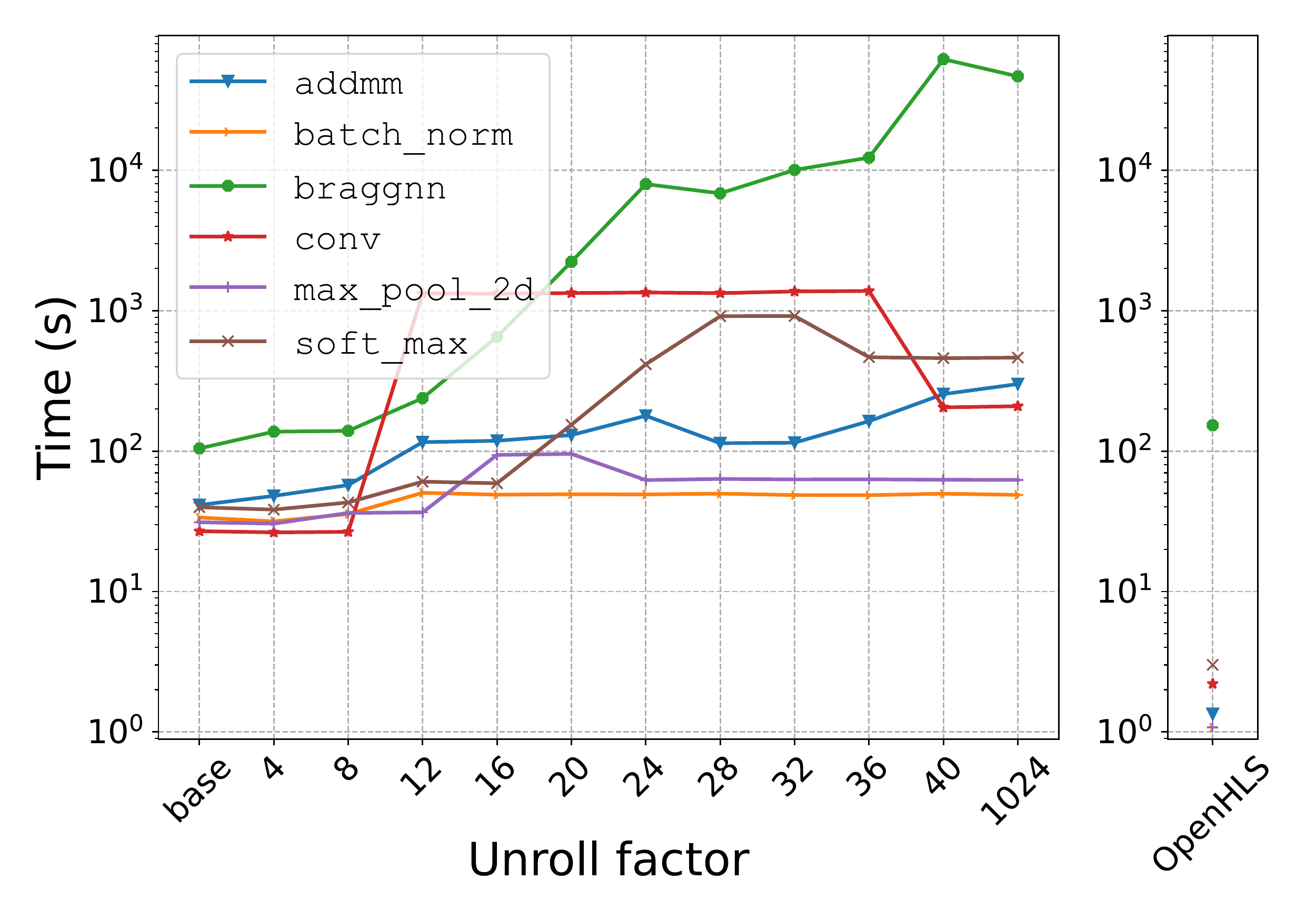}\caption{Vitis HLS vs. \texttt{OpenHLS} runtime vs. unroll factor, illustrating the large runtime cost incurred in using Vitis HLS to search over possible low-latency \texttt{BraggNN} designs.
\label{fig:Runtime-of-Vitisa}}
\end{figure}

Figure~\ref{fig:Resource-usage-and} shows Vitis HLS vs. \texttt{OpenHLS}
resource usage and latency vs. unroll factor and Figure~\ref{fig:Runtime-of-Vitisa}
shows the runtimes of Vitis HLS as function of increasing unroll factor.
We observe that while Vitis HLS end-to-end
latencies decrease with increased unroll factor, they never match that achieved
by \texttt{OpenHLS}. Even at an unroll factor of 1024 (which corresponds
to fully unrolled for all loop nests comprising these layer
types), Vitis HLS is only within 10$\times$ of \texttt{OpenHLS}.
We attribute this 
to Vitis HLS's inability
to pipeline effectively, due to its inability to eliminate memory dependencies,
either through \texttt{store}-\texttt{load} forwarding or further
array partitioning.
Conversely, \texttt{OpenHLS}'s ability to effectively perform \texttt{store}-\texttt{load}
forwarding is evident in the complete lack of BRAM usage: all weights
are kept on FFs or LUTs. While infeasible for larger designs
(which would be constrained by the number of available FFs), this unconstrained usage of FFs is acceptable for our use case.
The increasing latency (as a function of unroll factor) in the \texttt{max\_pool\_2d} case is due to Vitis HLS's failure to meet
timing, i.e., while the interval count decreases as a function of
unroll factor, the clock period increases.

\subsection{\texttt{BraggNN} case study\label{sec:BraggNN-case-study}}

\begin{figure}[tbh]
\centering{}\includegraphics[width=1\columnwidth,trim=0 8mm 0 7mm,clip]{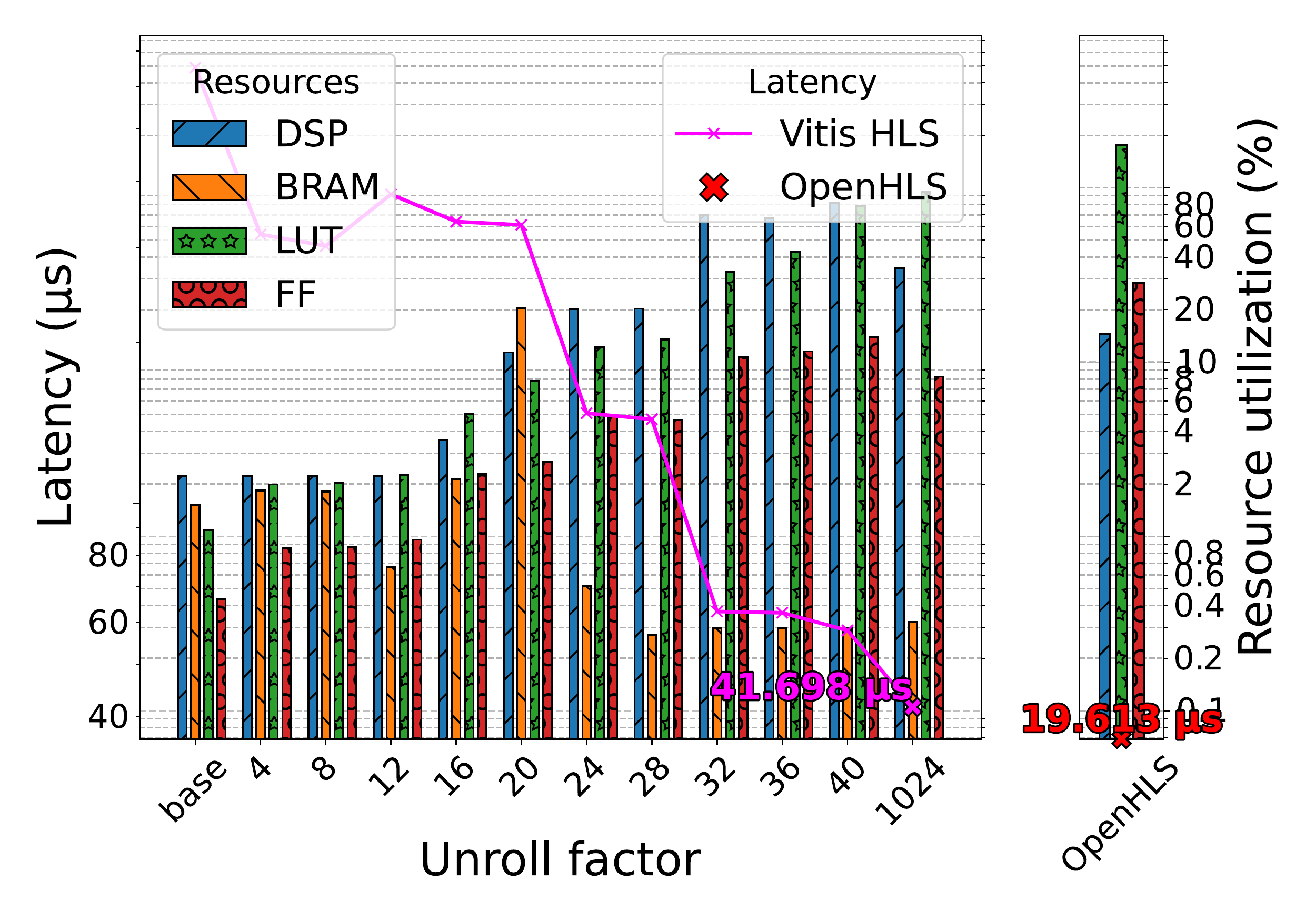}\caption{\texttt{BraggNN} Vitis HLS vs. \texttt{OpenHLS} resource usage and
latency vs. unroll factor (with both at half-precision) throughout the design space of possible low-latency designs.\texttt{\label{fig:braggnn}}}
\end{figure}

High-energy diffraction microscopy enables non-destructive
characterization for a broad class of single-crystal and polycrystalline
materials. 
A critical step in a typical HEDM experiment is an analysis to determine precise Bragg diffraction peak characteristics.
Peak characteristics are typically computed
by fitting the peaks to a probability distribution, e.g., Gaussian,
Lorentzian, Voigt, or Pseudo-Voigt. As noted in Section~\ref{sec:Introduction}, HEDM experiments can collect data
at more than 80 GB/s. These data rates, though more modest than at the LHC, 
merit exploring low latency approaches 
in order to enable experiment modalities that
depend on measurement-based feedback (i.e., experiment steering).
\begin{listing}
\begin{minted}[fontsize={\small},escapeinside={||},mathescape=true]{python}
BraggNN(|$s$|)(
  (cnn_layers_1): Conv2d(|$s \times 16 $|, kernel=3, stride=1)
  (nlb): NLB(
    (theta_layer): Conv2d(|$s \times 16 $|, |$s \times 8 $|, kernel=1, stride=1)
    (phi_layer): Conv2d(|$s \times 16 $|, |$s \times 8 $|, kernel=1, stride=1)
    (g_layer): Conv2d(|$s \times 16 $|, |$s \times 8 $|, kernel=1, stride=1)
    (out_cnn): Conv2d(|$s \times 8 $|, |$s \times 16 $|, kernel=1, stride=1)
    (soft): Softmax()
  )
  (cnn_layers_2): Sequential(
    (0): ReLU()
    (1): Conv2d(|$s \times 16 $|, |$s \times 8 $|, kernel=3, stride=1)
    (2): ReLU()
    (3): Conv2d(|$s \times 8 $|, |$s \times 2 $|, kernel=3, stride=1)
    (4): ReLU()
  )
  (dense_layers): Sequential(
    (0): Linear(in_features=|$s \times 50$|, out_features=|$s \times 16 $|)
    (1): ReLU()
    (2): Linear(in_features=|$s \times 16 $|, out_features=|$s \times 8 $|)
    (3): ReLU()
    (4): Linear(in_features=|$s \times 8 $|, out_features=|$s \times 4 $|)
    (5): ReLU()
    (6): Linear(in_features=|$s \times 4 $|, out_features=2)
    (7): ReLU()
  )
)
\end{minted}

\vspace{-1ex}

\caption{\texttt{BraggNN} model architecture for scaling factors \emph{s=1,2}.\label{lis:braggnn}}
\end{listing}

\texttt{BraggNN}~\cite{Liu:fs5198}, a DNN aimed at efficiently characterizing
Bragg diffraction peaks, achieves a throughput (via batch inference) of approximately 22
\textmu s/sample on a state-of-the-art GPU: a large speedup over classical pseudo-Voigt peak fitting methods, but still far
short of the 1 \textmu s/sample needed to handle 1 MHz sampling
rates. In addition, the 
data-center class GPU such as a NVIDIA V100 (or even a workstation class GPU such as a NVIDIA RTX 2080Ti) required to run the current  \texttt{BraggNN} implementation cannot be 
deployed at the edge, i.e., adjacent or
proximal to the high energy microscopy equipment. With the goal of reducing both per-sample time and deployment footprint,
we applied \texttt{OpenHLS} to the PyTorch representation of \texttt{BraggNN(\emph{s}=1)}\emph{
}(see Listing~\ref{lis:braggnn}) and achieved a RTL implementation
which synthesizes to a 1238 interval count design that places, routes,
and meets timing closure for a clock period of 10 ns (for a Xilinx
Alveo U280). The design consists of a three stage pipeline with the
longest stage measuring 480 intervals, for a throughput of 4.8 \textmu s/sample. See Figure~\ref{fig:braggnn} for a comparison with designs generated by Vitis HLS (using the same flow as in~\ref{sec:Evaluation}).

\begin{figure}[tbh]
\centering{}
\includegraphics[width=1\columnwidth,trim=5mm 4mm 4mm 3mm,clip]{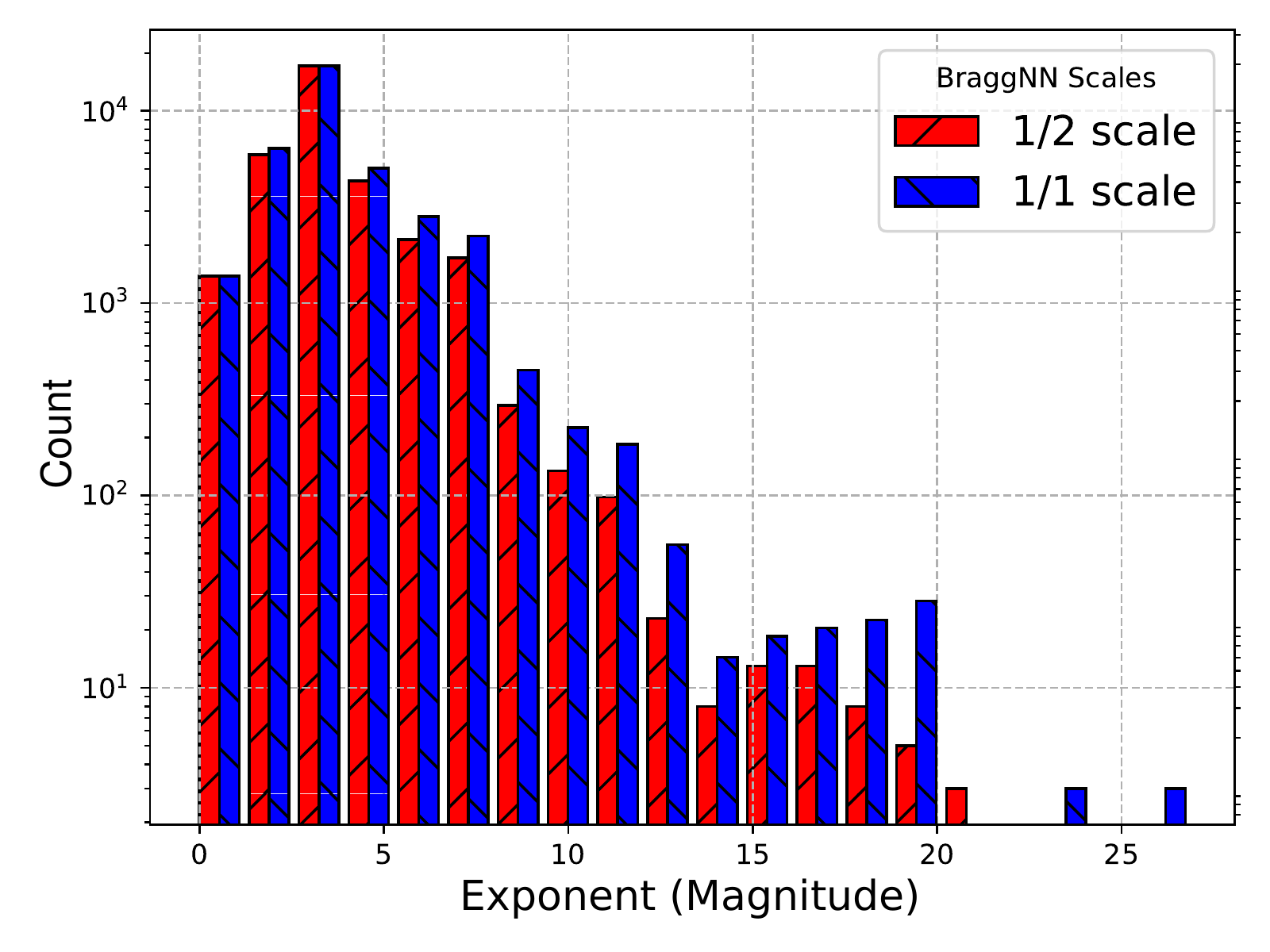}
\caption{\texttt{OpenHLS} weights exponent distribution, illustrating the narrow distribution of observed weight exponents thereby justifying reduced precision.\label{fig:OpenHLS-framework-overview.-1}}
\end{figure}

The most challenging aspect of implementing \texttt{BraggNN} was minimizing
latency while satisfying compute resource constraints (LUTs, DSPs,
BRAMs) and achieving routing closure, i.e., not exceeding available
routing resources and avoiding congestion. We made two design choices to reduce resource consumption. The first was to reduce the precision used for the floating-point operations,
from half precision to FloPoCo (5,4)-precision
(5-bit exponent, 4-bit mantissa), a choice justified
by examination of the distribution of the weights of the fully
trained \texttt{BraggNN} (see Figure~\ref{fig:OpenHLS-framework-overview.-1}).

Reducing the precision enabled the second design choice, to eliminate
BRAMs from the design, since, at the lower precision, all weights
can be represented as registered constants. The reduced precision
also drove the Vivado synthesizer to infer implementations of the
floating-point operations that make no use of DSPs,
likely becaue
the DSP48 hardware block includes a 18-bit by 25-bit signed multiplier
and a 48-bit adder~\cite{guideultrascale}, neither of which neatly
divides the bit width of FloPoCo (5,4)-precision cores. 
(The actual width for FloPoCo (5,4)-precision is 12 bits: 1 extra bit is needed for the sign and 2 for handling of exceptional conditions.)

\begin{figure*}[tbh]
\centering{}\subfloat[\texttt{BraggNN} fails to achieve routing closure without placement
and routing constraints and launch and latch registers.\label{fig:OpenHLS-framework-overview.-2-1}]{\centering{}\includegraphics[width=1\columnwidth]{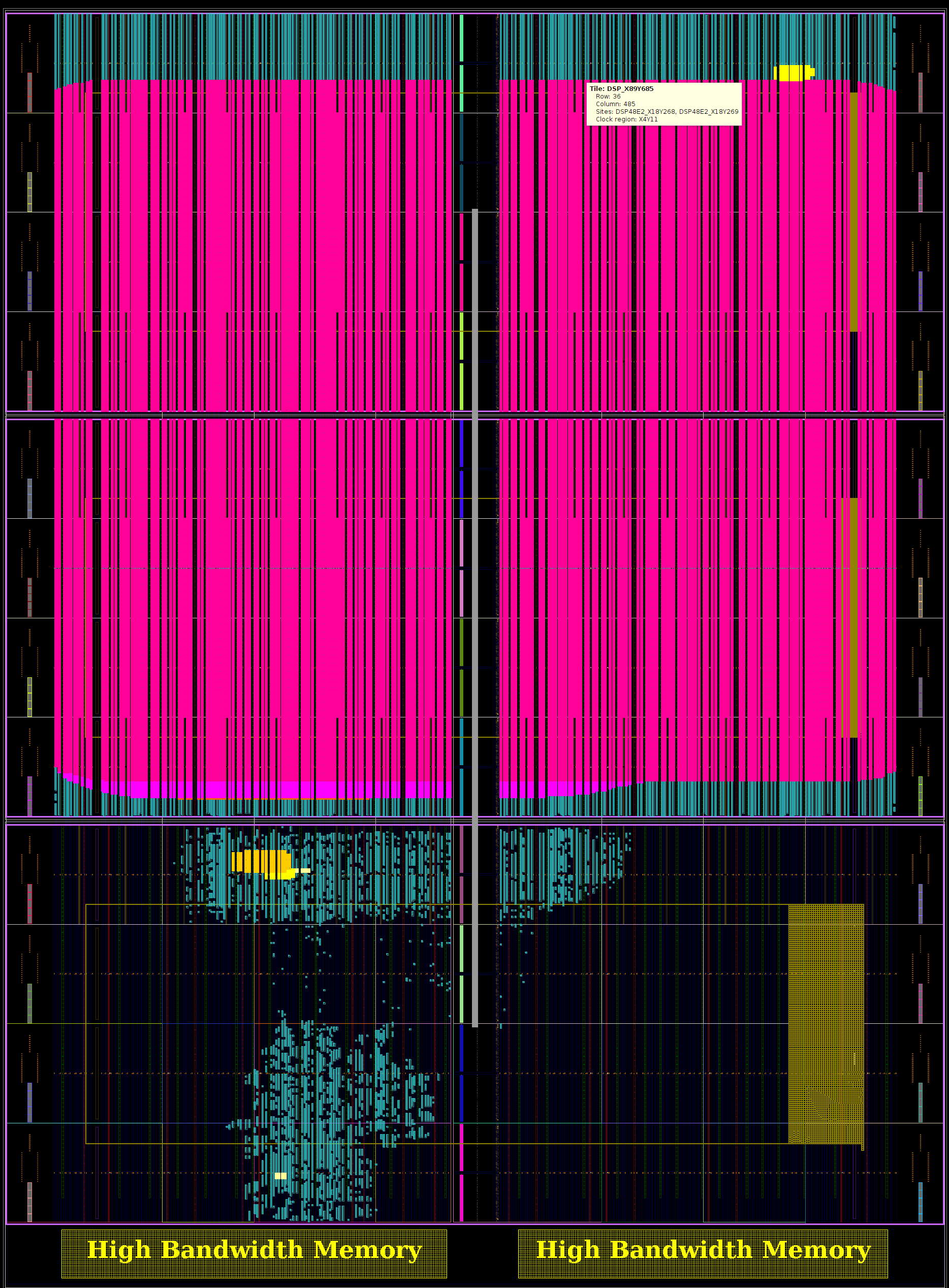}}\hfill{}\subfloat[\texttt{BraggNN} achieves routing closure with use of per SLR placement
and routing constraints (\texttt{pblock\_1}, \texttt{pblock\_2}, \texttt{pblock\_3})
and launch and latch registers (not highlighted).\label{fig:OpenHLS-framework-overview.-2-1-1}]{\centering{}\includegraphics[width=1\columnwidth]{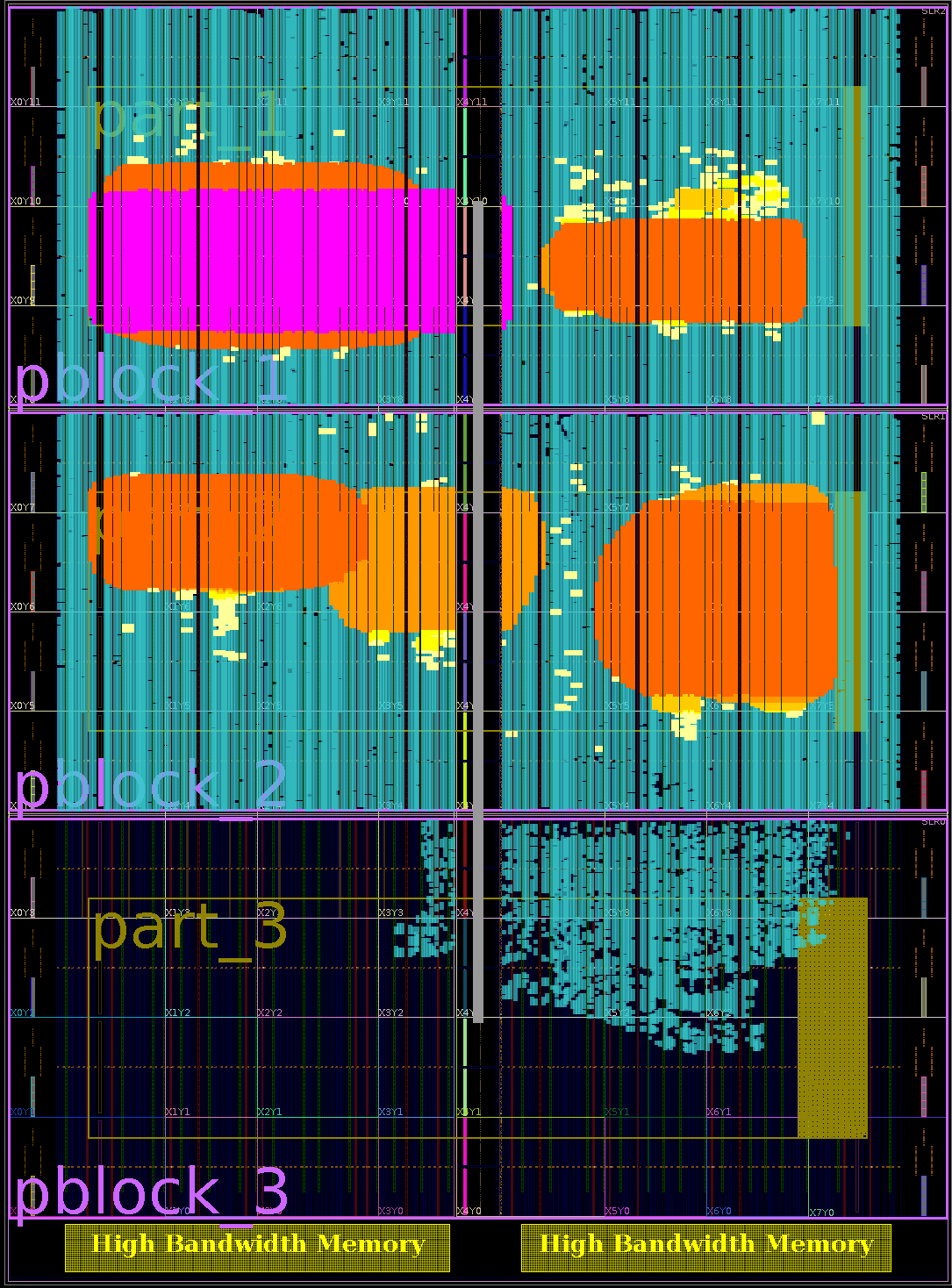}}\caption{Congestion maps for \texttt{BraggNN} on a Xilinx Alveo U280. \textcolor{magenta}{Magenta}
indicates areas of high congestion.\label{fig:OpenHLS-framework-overview.-2}}
\end{figure*}

Achieving routing closure was difficult due to the nature of
the Xilinx's UltraScale architecture, of which the Alveo U280 is an
instance. The UltraScale architecture achieves its scale through Stacked
Silicon Interconnect (SSI) technology~\cite{leibson2013xilinx},
which implies multiple distinct FPGA dies, called Super Logic Regions
(SLRs), on the same chip, connected by interposers. Adjacent SLRs
communicate with each other over a limited set of Super Long Lines
(SLLs), which determine the maximum bus width that spans two
SLRs. On the Alveo U280 there are exactly 23,040 SLLs available between
adjacent SLRs and at (5,4)-precision \texttt{BraggNN}(\emph{s}\/=1)
needs 23,328 SLLs between SLR2 and SLR1.
[We route from SLR2 to SLR1 the outputs of \texttt{cnn\_layers\_1} (1$\times$16$\times$9$\times$9$\times$12
wires) and \texttt{soft(theta\_layer}$\times$ \texttt{phi\_layer)}$\times$\texttt{g\_layer} (1$\times$8$\times$9$\times$9$\times$12 wires).] Thus, we further reduced the precision to (5,3). Finally,
since multiple dies constitute independent clock domains, the SLLs
that cross SLRS are sensitive to hold time violations due to the higher
multi-die variability~\cite{rapidwright}. This multi-die variability
leads to high congestion if not addressed. Thus, routing across SLRs
needs to be handled manually, using placement and routing constraints
for logic in each SLR and the addition of so-called ``launch'' and
``latch'' registers in each SLR. 
Figure~\ref{fig:OpenHLS-framework-overview.-2} illustrates the effect of using launch and latch registers as well as placement and routing constraints.

Thus, these design choices (in combination
with compiler level optimizations performed by \texttt{OpenHLS})
plus careful management of routing constraints enable us to lower,
compile, synthesize, place, and route \texttt{BraggNN}(\emph{s}\/=1) to Xilinx's
Alveo U280 at a throughput of 4.8 \textmu s/sample: \textasciitilde 5$\times$
higher latency than the target 1 \textmu s/sample, but a \textasciitilde 4$\times$
improvement over the PyTorch GPU implementation. 

\section{Conclusion\label{sec:Conclusion}}

We have presented \texttt{OpenHLS}, an MLIR-based HLS compilation
framework that supports translating DNN models to RTL without the use of
commercial HLS tools. The \texttt{OpenHLS} end-to-end
compilation pipeline provides a PyTorch front-end and Verilog
emission back-end. An extensible Python
intermediate layer supports use-case-specific optimizations (e.g., \texttt{store}-\texttt{load} forwarding) that are
not possible otherwise. Experimental results demonstrate that \texttt{OpenHLS}
outperforms, in terms of end-to-end latency, Vitis HLS on a range
of DNN layer types and on a case-study DNN.


We note three directions for future work, primarily with respect to scheduling:
(1) Better integration between the Python layer and MLIR: it is preferable that the transformations on the Python representation could make use of various MLIR facilities, such as affine analysis, for the purposes of exploring loop transformations that improve latency;
(2) Expanding the set of scheduling algorithms available: for example, resource aware scheduling~\cite{10.1145/3174243.3174268}; and
(3) Integration of scheduling-aware placement and vice-versa (placement-aware scheduling): currently \texttt{OpenHLS} can be used to inform placement but does not explicitly emit placement constraints (see Section~\ref{sec:BraggNN-case-study}); a more precise approach, such as in~\cite{10.1145/3431920.3439289}, would potentially enable better pipelining and thus higher throughput.

\bibliographystyle{ACM-Reference-Format}
\bibliography{bragghls}


\begin{thebibliography}{49}


\ifx \showCODEN    \undefined \def \showCODEN     #1{\unskip}     \fi
\ifx \showDOI      \undefined \def \showDOI       #1{#1}\fi
\ifx \showISBNx    \undefined \def \showISBNx     #1{\unskip}     \fi
\ifx \showISBNxiii \undefined \def \showISBNxiii  #1{\unskip}     \fi
\ifx \showISSN     \undefined \def \showISSN      #1{\unskip}     \fi
\ifx \showLCCN     \undefined \def \showLCCN      #1{\unskip}     \fi
\ifx \shownote     \undefined \def \shownote      #1{#1}          \fi
\ifx \showarticletitle \undefined \def \showarticletitle #1{#1}   \fi
\ifx \showURL      \undefined \def \showURL       {\relax}        \fi
\providecommand\bibfield[2]{#2}
\providecommand\bibinfo[2]{#2}
\providecommand\natexlab[1]{#1}
\providecommand\showeprint[2][]{arXiv:#2}

\bibitem[rap({[n.\,d.]})]%
        {rapidwright}
\bibinfo{title}{Create placed and routed DCP to cross SLR}.
\newblock
  \bibinfo{howpublished}{\url{https://www.rapidwright.io/docs/SLR_Crosser_DCP_Creator_Tutorial.html}}.
\newblock
\newblock
\shownote{Accessed: 2022-10-15}.


\bibitem[gui(2021)]%
        {guideultrascale}
 \bibinfo{year}{2021}\natexlab{}.
\newblock \bibinfo{booktitle}{\emph{UltraScale Architecture DSP Slice}}.
\newblock \bibinfo{type}{{T}echnical {R}eport}. \bibinfo{institution}{XiLinx}.
\newblock
\urldef\tempurl%
\url{https://docs.xilinx.com/v/u/en-US/ug579-ultrascale-dsp}
\showURL{%
\tempurl}


\bibitem[Aaij et~al\mbox{.}(2020)]%
        {aaij2020allen}
\bibfield{author}{\bibinfo{person}{Roel Aaij} {et~al\mbox{.}}}
  \bibinfo{year}{2020}\natexlab{}.
\newblock \showarticletitle{Allen: A high-level trigger on GPUs for LHCb}.
\newblock \bibinfo{journal}{\emph{Computing and Software for Big Science}}
  \bibinfo{volume}{4}, \bibinfo{number}{1} (\bibinfo{year}{2020}),
  \bibinfo{pages}{1--11}.
\newblock


\bibitem[Abadi et~al\mbox{.}(2016)]%
        {https://doi.org/10.48550/arxiv.1603.04467}
\bibfield{author}{\bibinfo{person}{Mart\'{i}n Abadi} {et~al\mbox{.}}}
\newblock \bibinfo{title}{TensorFlow: Large-Scale Machine Learning on
  Heterogeneous Distributed Systems}.
\newblock
\newblock
\urldef\tempurl%
\url{https://doi.org/10.48550/ARXIV.1603.04467}
\showDOI{\tempurl}


\bibitem[Alzubaidi et~al\mbox{.}(2021)]%
        {alzubaidi2021review}
\bibfield{author}{\bibinfo{person}{Laith Alzubaidi} {et~al\mbox{.}}}
  \bibinfo{year}{2021}\natexlab{}.
\newblock \showarticletitle{Review of deep learning: Concepts, CNN
  architectures, challenges, applications, future directions}.
\newblock \bibinfo{journal}{\emph{Journal of Big Data}} \bibinfo{volume}{8},
  \bibinfo{number}{1} (\bibinfo{year}{2021}), \bibinfo{pages}{1--74}.
\newblock


\bibitem[Ashari et~al\mbox{.}(2015)]%
        {10.1145/2858788.2688521}
\bibfield{author}{\bibinfo{person}{Arash Ashari} {et~al\mbox{.}}}
  \bibinfo{year}{2015}\natexlab{}.
\newblock \showarticletitle{On Optimizing Machine Learning Workloads via Kernel
  Fusion}.
\newblock  \bibinfo{volume}{50}, \bibinfo{number}{8} (\bibinfo{year}{2015}),
  \bibinfo{pages}{173--182}.
\newblock
\showISSN{0362-1340}
\urldef\tempurl%
\url{https://doi.org/10.1145/2858788.2688521}
\showDOI{\tempurl}


\bibitem[Baruch(1996)]%
        {baruch1996scheduling}
\bibfield{author}{\bibinfo{person}{Zoltan Baruch}.}
  \bibinfo{year}{1996}\natexlab{}.
\newblock \showarticletitle{Scheduling algorithms for high-level synthesis}.
\newblock \bibinfo{journal}{\emph{ACAM Scientific Journal}}
  \bibinfo{volume}{5}, \bibinfo{number}{1-2} (\bibinfo{year}{1996}),
  \bibinfo{pages}{48--57}.
\newblock


\bibitem[Bohm~Agostini et~al\mbox{.}(2022)]%
        {9786533}
\bibfield{author}{\bibinfo{person}{Nicolas Bohm~Agostini} {et~al\mbox{.}}}
  \bibinfo{year}{2022}\natexlab{}.
\newblock \showarticletitle{Bridging Python to Silicon: The SODA Toolchain}.
\newblock \bibinfo{journal}{\emph{IEEE Micro}} (\bibinfo{year}{2022}).
\newblock
\urldef\tempurl%
\url{https://doi.org/10.1109/MM.2022.3178580}
\showDOI{\tempurl}


\bibitem[Bondhugula(2020)]%
        {polyhedral-mlir}
\bibfield{author}{\bibinfo{person}{Uday Bondhugula}.}
\newblock \bibinfo{title}{Polyhedral compilation opportunities in MLIR}.
\newblock
  \bibinfo{howpublished}{\url{https://acohen.gitlabpages.inria.fr/impact/impact2020/slides/IMPACT_2020_keynote.pdf}}.
\newblock


\bibitem[Canis et~al\mbox{.}(2013)]%
        {10.1145/2514740}
\bibfield{author}{\bibinfo{person}{Andrew Canis} {et~al\mbox{.}}}
  \bibinfo{year}{2013}\natexlab{}.
\newblock \showarticletitle{LegUp: An Open-Source High-Level Synthesis Tool for
  FPGA-Based Processor/Accelerator Systems}.
\newblock \bibinfo{journal}{\emph{ACM Trans. Embed. Comput. Syst.}}
  \bibinfo{volume}{13}, \bibinfo{number}{2}, Article \bibinfo{articleno}{24}
  (\bibinfo{year}{2013}).
\newblock
\showISSN{1539-9087}
\urldef\tempurl%
\url{https://doi.org/10.1145/2514740}
\showDOI{\tempurl}


\bibitem[Chen et~al\mbox{.}(2015)]%
        {https://doi.org/10.48550/arxiv.1512.01274}
\bibfield{author}{\bibinfo{person}{Tianqi Chen} {et~al\mbox{.}}}
\newblock \bibinfo{title}{MXNet: A Flexible and Efficient Machine Learning
  Library for Heterogeneous Distributed Systems}.
\newblock
\newblock
\urldef\tempurl%
\url{https://doi.org/10.48550/ARXIV.1512.01274}
\showDOI{\tempurl}


\bibitem[Chen et~al\mbox{.}(2018)]%
        {chen2018tvm}
\bibfield{author}{\bibinfo{person}{Tianqi Chen} {et~al\mbox{.}}}
  \bibinfo{year}{2018}\natexlab{}.
\newblock \showarticletitle{TVM: An automated end-to-end optimizing compiler
  for deep learning}. In \bibinfo{booktitle}{\emph{13th USENIX Symp.\ Operating
  Systems Design \& Impl.}} \bibinfo{pages}{578--594}.
\newblock


\bibitem[Chen et~al\mbox{.}(2016)]%
        {https://doi.org/10.48550/arxiv.1604.06174}
\bibfield{author}{\bibinfo{person}{Tianqi Chen}, \bibinfo{person}{Bing Xu},
  \bibinfo{person}{Chiyuan Zhang}, {and} \bibinfo{person}{Carlos Guestrin}.}
\newblock \bibinfo{title}{Training Deep Nets with Sublinear Memory Cost}.
\newblock
\newblock
\urldef\tempurl%
\url{https://doi.org/10.48550/ARXIV.1604.06174}
\showDOI{\tempurl}


\bibitem[Collaboration(2020)]%
        {LHCB-FIGURE-2020-018}
\bibfield{author}{\bibinfo{person}{LHCb Collaboration}.}
  \bibinfo{year}{2020}\natexlab{}.
\newblock \bibinfo{booktitle}{\emph{Comparison of particle selection algorithms
  for the {LHCb} Upgrade}}.
\newblock \bibinfo{type}{{T}echnical {R}eport}.
\newblock
\urldef\tempurl%
\url{https://cds.cern.ch/record/2746789}
\showURL{%
\tempurl}


\bibitem[Dai et~al\mbox{.}(2018)]%
        {10.1145/3174243.3174268}
\bibfield{author}{\bibinfo{person}{Steve Dai}, \bibinfo{person}{Gai Liu}, {and}
  \bibinfo{person}{Zhiru Zhang}.} \bibinfo{year}{2018}\natexlab{}.
\newblock \showarticletitle{A Scalable Approach to Exact Resource-Constrained
  Scheduling Based on a Joint SDC and SAT Formulation}. In
  \bibinfo{booktitle}{\emph{ACM/SIGDA Intl Symposium on Field-Programmable Gate
  Arrays}}. \bibinfo{pages}{137--146}.
\newblock
\showISBNx{9781450356145}


\bibitem[de~Dinechin(2019)]%
        {8877424}
\bibfield{author}{\bibinfo{person}{Florent de Dinechin}.}
  \bibinfo{year}{2019}\natexlab{}.
\newblock \showarticletitle{Reflections on 10 Years of FloPoCo}. In
  \bibinfo{booktitle}{\emph{IEEE 26th Symposium on Computer Arithmetic}}.
  \bibinfo{pages}{187--189}.
\newblock


\bibitem[Duarte et~al\mbox{.}(2018)]%
        {Duarte_2018}
\bibfield{author}{\bibinfo{person}{J. Duarte} {et~al\mbox{.}}}
  \bibinfo{year}{2018}\natexlab{}.
\newblock \showarticletitle{Fast inference of deep neural networks in {FPGAs}
  for particle physics}.
\newblock \bibinfo{journal}{\emph{Journal of Instrumentation}}
  \bibinfo{volume}{13}, \bibinfo{number}{07} (\bibinfo{year}{2018}),
  \bibinfo{pages}{P07027--P07027}.
\newblock


\bibitem[Ferrandi et~al\mbox{.}(2021)]%
        {ferrandi2021bambu}
\bibfield{author}{\bibinfo{person}{Fabrizio Ferrandi} {et~al\mbox{.}}}
  \bibinfo{year}{2021}\natexlab{}.
\newblock \showarticletitle{Bambu: an Open-Source Research Framework for the
  High-Level Synthesis of Complex Applications}. In
  \bibinfo{booktitle}{\emph{58th ACM/IEEE Design Automation Conference}}.
  \bibinfo{publisher}{{IEEE}}, \bibinfo{pages}{1327--1330}.
\newblock


\bibitem[Gligorov(2015)]%
        {pmlr-v42-glig14}
\bibfield{author}{\bibinfo{person}{Vladimir Gligorov}.}
  \bibinfo{year}{2015}\natexlab{}.
\newblock \showarticletitle{Real-time data analysis at the {LHC}: present and
  future}. In \bibinfo{booktitle}{\emph{NIPS Workshop on High-energy Physics
  and Machine Learning}}, Vol.~\bibinfo{volume}{42}. \bibinfo{pages}{1--18}.
\newblock


\bibitem[Gligorov and Williams(2013)]%
        {Gligorov_2013}
\bibfield{author}{\bibinfo{person}{V~V Gligorov} {and} \bibinfo{person}{M
  Williams}.} \bibinfo{year}{2013}\natexlab{}.
\newblock \showarticletitle{Efficient, reliable and fast high-level triggering
  using a bonsai boosted decision tree}.
\newblock \bibinfo{journal}{\emph{J.\ Instrumentation}} \bibinfo{volume}{8},
  \bibinfo{number}{02} (\bibinfo{year}{2013}).
\newblock


\bibitem[Grainge et~al\mbox{.}(2017)]%
        {grainge2017square}
\bibfield{author}{\bibinfo{person}{Keith Grainge} {et~al\mbox{.}}}
  \bibinfo{year}{2017}\natexlab{}.
\newblock \showarticletitle{Square Kilometre Array: The radio telescope of the
  XXI century}.
\newblock \bibinfo{journal}{\emph{Astronomy reports}} \bibinfo{volume}{61},
  \bibinfo{number}{4} (\bibinfo{year}{2017}), \bibinfo{pages}{288--296}.
\newblock


\bibitem[Guo et~al\mbox{.}(2021)]%
        {10.1145/3431920.3439289}
\bibfield{author}{\bibinfo{person}{Licheng Guo} {et~al\mbox{.}}}
  \bibinfo{year}{2021}\natexlab{}.
\newblock \showarticletitle{AutoBridge: Coupling Coarse-Grained Floorplanning
  and Pipelining for High-Frequency HLS Design on Multi-Die FPGAs}. In
  \bibinfo{booktitle}{\emph{ACM/SIGDA International Symposium on
  Field-Programmable Gate Arrays}}. \bibinfo{pages}{81–92}.
\newblock
\showISBNx{9781450382182}


\bibitem[Hammer et~al\mbox{.}(2021)]%
        {Hammer_2021}
\bibfield{author}{\bibinfo{person}{M. Hammer}, \bibinfo{person}{K. Yoshii},
  {and} \bibinfo{person}{A. Miceli}.} \bibinfo{year}{2021}\natexlab{}.
\newblock \showarticletitle{Strategies for on-chip digital data compression for
  X-ray pixel detectors}.
\newblock \bibinfo{journal}{\emph{Journal of Instrumentation}}
  \bibinfo{volume}{16}, \bibinfo{number}{01} (\bibinfo{year}{2021}),
  \bibinfo{pages}{P01025--P01025}.
\newblock
\urldef\tempurl%
\url{https://doi.org/10.1088/1748-0221/16/01/p01025}
\showDOI{\tempurl}


\bibitem[Hattori et~al\mbox{.}(2022)]%
        {https://doi.org/10.48550/arxiv.2203.08402}
\bibfield{author}{\bibinfo{person}{Momoko Hattori}, \bibinfo{person}{Naoki
  Kobayashi}, {and} \bibinfo{person}{Ryosuke Sato}.}
\newblock \bibinfo{title}{Gradual Tensor Shape Checking}.
\newblock
\newblock
\urldef\tempurl%
\url{https://doi.org/10.48550/ARXIV.2203.08402}
\showDOI{\tempurl}


\bibitem[Ioffe and Szegedy(2015)]%
        {https://doi.org/10.48550/arxiv.1502.03167}
\bibfield{author}{\bibinfo{person}{Sergey Ioffe} {and}
  \bibinfo{person}{Christian Szegedy}.}
\newblock \bibinfo{title}{Batch Normalization: Accelerating Deep Network
  Training by Reducing Internal Covariate Shift}.
\newblock
\newblock
\urldef\tempurl%
\url{https://doi.org/10.48550/ARXIV.1502.03167}
\showDOI{\tempurl}


\bibitem[Lattner et~al\mbox{.}(2020)]%
        {https://doi.org/10.48550/arxiv.2002.11054}
\bibfield{author}{\bibinfo{person}{Chris Lattner} {et~al\mbox{.}}}
\newblock \bibinfo{title}{MLIR: A compiler infrastructure for the end of
  Moore's Law}.
\newblock
\newblock
\urldef\tempurl%
\url{https://doi.org/10.48550/ARXIV.2002.11054}
\showDOI{\tempurl}


\bibitem[Leibson et~al\mbox{.}(2013)]%
        {leibson2013xilinx}
\bibfield{author}{\bibinfo{person}{Steve Leibson} {et~al\mbox{.}}}
  \bibinfo{year}{2013}\natexlab{}.
\newblock \showarticletitle{Xilinx ultrascale: The next-generation architecture
  for your next-generation architecture}.
\newblock \bibinfo{journal}{\emph{Xilinx White Paper WP435}}
  \bibinfo{volume}{143} (\bibinfo{year}{2013}).
\newblock


\bibitem[Liu et~al\mbox{.}(2018)]%
        {https://doi.org/10.48550/arxiv.1809.02697}
\bibfield{author}{\bibinfo{person}{Yizhi Liu} {et~al\mbox{.}}}
  \bibinfo{year}{2018}\natexlab{}.
\newblock \showarticletitle{Optimizing CNN Model Inference on CPUs}.
\newblock  (\bibinfo{year}{2018}).
\newblock
\urldef\tempurl%
\url{https://doi.org/10.48550/ARXIV.1809.02697}
\showDOI{\tempurl}


\bibitem[Liu et~al\mbox{.}(2022a)]%
        {liu2022exploring}
\bibfield{author}{\bibinfo{person}{Yongtao Liu} {et~al\mbox{.}}}
  \bibinfo{year}{2022}\natexlab{a}.
\newblock \showarticletitle{Exploring physics of ferroelectric domain walls in
  real time: Deep learning enabled scanning probe microscopy}.
\newblock \bibinfo{journal}{\emph{Advanced Science}} (\bibinfo{year}{2022}).
\newblock


\bibitem[Liu et~al\mbox{.}(2019)]%
        {liu2019deep}
\bibfield{author}{\bibinfo{person}{Zhengchun Liu} {et~al\mbox{.}}}
  \bibinfo{year}{2019}\natexlab{}.
\newblock \showarticletitle{Deep learning accelerated light source
  experiments}. In \bibinfo{booktitle}{\emph{IEEE/ACM 3rd Workshop on Deep
  Learning on Supercomputers}}. IEEE, \bibinfo{pages}{20--28}.
\newblock


\bibitem[Liu et~al\mbox{.}(2022b)]%
        {Liu:fs5198}
\bibfield{author}{\bibinfo{person}{Zhengchun Liu} {et~al\mbox{.}}}
  \bibinfo{year}{2022}\natexlab{b}.
\newblock \showarticletitle{{{\it BraggNN}: fast X-ray Bragg peak analysis
  using deep learning}}.
\newblock \bibinfo{journal}{\emph{IUCrJ}} \bibinfo{volume}{9},
  \bibinfo{number}{1} (\bibinfo{year}{2022}), \bibinfo{pages}{104--113}.
\newblock


\bibitem[Maleki et~al\mbox{.}(2011)]%
        {maleki2011evaluation}
\bibfield{author}{\bibinfo{person}{Saeed Maleki} {et~al\mbox{.}}}
  \bibinfo{year}{2011}\natexlab{}.
\newblock \showarticletitle{An evaluation of vectorizing compilers}. In
  \bibinfo{booktitle}{\emph{International Conference on Parallel Architectures
  and Compilation Techniques}}. IEEE, \bibinfo{pages}{372--382}.
\newblock


\bibitem[McMullin et~al\mbox{.}(2022)]%
        {mcmullin2022square}
\bibfield{author}{\bibinfo{person}{J McMullin} {et~al\mbox{.}}}
  \bibinfo{year}{2022}\natexlab{}.
\newblock \showarticletitle{The Square Kilometre Array project update}. In
  \bibinfo{booktitle}{\emph{Ground-based and Airborne Telescopes IX}},
  Vol.~\bibinfo{volume}{12182}. SPIE, \bibinfo{pages}{263--271}.
\newblock


\bibitem[Nane et~al\mbox{.}(2016)]%
        {7368920}
\bibfield{author}{\bibinfo{person}{Razvan Nane} {et~al\mbox{.}}}
  \bibinfo{year}{2016}\natexlab{}.
\newblock \showarticletitle{A Survey and Evaluation of FPGA High-Level
  Synthesis Tools}.
\newblock \bibinfo{journal}{\emph{IEEE Transactions on Computer-Aided Design of
  Integrated Circuits and Systems}} \bibinfo{volume}{35}, \bibinfo{number}{10}
  (\bibinfo{year}{2016}), \bibinfo{pages}{1591--1604}.
\newblock
\urldef\tempurl%
\url{https://doi.org/10.1109/TCAD.2015.2513673}
\showDOI{\tempurl}


\bibitem[Oppermann(2019)]%
        {tuprints9272}
\bibfield{author}{\bibinfo{person}{Julian Oppermann}.}
  \bibinfo{year}{2019}\natexlab{}.
\newblock \emph{\bibinfo{title}{Advances in ILP-based Modulo Scheduling for
  High-Level Synthesis}}.
\newblock \bibinfo{thesistype}{Ph.\,D. Dissertation}.
  \bibinfo{school}{Technische Universit{\"a}t}, \bibinfo{address}{Darmstadt}.
\newblock
\urldef\tempurl%
\url{http://tuprints.ulb.tu-darmstadt.de/9272/}
\showURL{%
\tempurl}


\bibitem[Oppermann et~al\mbox{.}(2022)]%
        {oppermann2022eurollvm}
\bibfield{author}{\bibinfo{person}{Julian Oppermann} {et~al\mbox{.}}}
  \bibinfo{year}{2022}\natexlab{}.
\newblock \showarticletitle{How to make hardware with maths: An introduction to
  {CIRCT}'s scheduling infrastructure}. In \bibinfo{booktitle}{\emph{European
  {LLVM} {Developers}' {Meeting}}}.
\newblock


\bibitem[Paszke et~al\mbox{.}(2017)]%
        {paszke2017automatic}
\bibfield{author}{\bibinfo{person}{Adam Paszke} {et~al\mbox{.}}}
  \bibinfo{year}{2017}\natexlab{}.
\newblock \showarticletitle{Automatic differentiation in {PyTorch}}. In
  \bibinfo{booktitle}{\emph{31st Conference on Neural Information Processing
  Systems}}.
\newblock


\bibitem[Patton et~al\mbox{.}(2018)]%
        {patton2018167}
\bibfield{author}{\bibinfo{person}{Robert~M Patton} {et~al\mbox{.}}}
  \bibinfo{year}{2018}\natexlab{}.
\newblock \showarticletitle{167-{P}flops deep learning for electron microscopy:
  From learning physics to atomic manipulation}. In
  \bibinfo{booktitle}{\emph{SC'18}}. IEEE, \bibinfo{pages}{638--648}.
\newblock


\bibitem[Rajopadhye(2002)]%
        {rajopadhye2002dependence}
\bibfield{author}{\bibinfo{person}{Sanjay~V Rajopadhye}.}
  \bibinfo{year}{2002}\natexlab{}.
\newblock \showarticletitle{Dependence Analysis and Parallelizing
  Transformations}.
\newblock In \bibinfo{booktitle}{\emph{The Compiler Design Handbook}}.
\newblock


\bibitem[Rausch et~al\mbox{.}(2022)]%
        {daceml}
\bibfield{author}{\bibinfo{person}{Oliver Rausch} {et~al\mbox{.}}}
  \bibinfo{year}{2022}\natexlab{}.
\newblock \showarticletitle{{DaCeML}: A Data-Centric Optimization Framework for
  Machine Learning}. In \bibinfo{booktitle}{\emph{36th ACM International
  Conference on Supercomputing}}.
\newblock


\bibitem[Rosser(2018)]%
        {rosser2018cocotb}
\bibfield{author}{\bibinfo{person}{Benjamin~John Rosser}.}
\newblock \bibinfo{title}{Cocotb: a Python-based digital logic verification
  framework}.
\newblock
\newblock
\newblock
\shownote{\url{https://docs.cocotb.org}}.


\bibitem[Rotem et~al\mbox{.}(2018)]%
        {https://doi.org/10.48550/arxiv.1805.00907}
\bibfield{author}{\bibinfo{person}{Nadav Rotem} {et~al\mbox{.}}}
\newblock \bibinfo{title}{Glow: Graph Lowering Compiler Techniques for Neural
  Networks}.
\newblock
\newblock
\urldef\tempurl%
\url{https://doi.org/10.48550/ARXIV.1805.00907}
\showDOI{\tempurl}


\bibitem[Sharma et~al\mbox{.}(2016)]%
        {7783720}
\bibfield{author}{\bibinfo{person}{Hardik Sharma} {et~al\mbox{.}}}
  \bibinfo{year}{2016}\natexlab{}.
\newblock \showarticletitle{From high-level deep neural models to FPGAs}. In
  \bibinfo{booktitle}{\emph{49th Annual IEEE/ACM International Symposium on
  Microarchitecture}}. \bibinfo{pages}{1--12}.
\newblock


\bibitem[Silva and Elangovan(2021)]%
        {torch-mlir}
\bibfield{author}{\bibinfo{person}{Sean Silva} {and} \bibinfo{person}{Anush
  Elangovan}.}
\newblock \bibinfo{title}{{Torch-MLIR}}.
\newblock
  \bibinfo{howpublished}{\url{https://mlir.llvm.org/OpenMeetings/2021-10-07-The-Torch-MLIR-project.pdf}}.
\newblock


\bibitem[Takamaeda-Yamazaki(2015)]%
        {takamaeda2015pyverilog}
\bibfield{author}{\bibinfo{person}{Shinya Takamaeda-Yamazaki}.}
  \bibinfo{year}{2015}\natexlab{}.
\newblock \showarticletitle{Pyverilog: A Python-based hardware design
  processing toolkit for Verilog HDL}. In
  \bibinfo{booktitle}{\emph{International Symposium on Applied Reconfigurable
  Computing}}. Springer, \bibinfo{pages}{451--460}.
\newblock


\bibitem[Williams({[n.\,d.]})]%
        {williamsicarus}
\bibfield{author}{\bibinfo{person}{Stephen Williams}.}
\newblock \bibinfo{title}{Icarus Verilog, 1998--2020}.
\newblock
\newblock
\newblock
\shownote{\url{http://iverilog.icarus.com}}.


\bibitem[Ye et~al\mbox{.}(2022)]%
        {yehpca2022scalehls}
\bibfield{author}{\bibinfo{person}{Hanchen Ye} {et~al\mbox{.}}}
  \bibinfo{year}{2022}\natexlab{}.
\newblock \showarticletitle{ScaleHLS: A New Scalable High-Level Synthesis
  Framework on Multi-Level Intermediate Representation}. In
  \bibinfo{booktitle}{\emph{IEEE International Symposium on High-Performance
  Computer Architecture}}.
\newblock


\bibitem[Zhang et~al\mbox{.}(2008)]%
        {Zhang2008}
\bibfield{author}{\bibinfo{person}{Zhiru Zhang} {et~al\mbox{.}}}
  \bibinfo{year}{2008}\natexlab{}.
\newblock \showarticletitle{AutoPilot: A Platform-Based ESL Synthesis System}.
\newblock In \bibinfo{booktitle}{\emph{High-Level Synthesis}}.
  \bibinfo{publisher}{Springer Netherlands}, \bibinfo{address}{Dordrecht},
  \bibinfo{pages}{99--112}.
\newblock
\showISBNx{978-1-4020-8588-8}


\bibitem[Zheng et~al\mbox{.}(2022)]%
        {9664259}
\bibfield{author}{\bibinfo{person}{S. Zheng} {et~al\mbox{.}}}
  \bibinfo{year}{2022}\natexlab{}.
\newblock \showarticletitle{NeoFlow: A Flexible Framework for Enabling
  Efficient Compilation for High Performance DNN Training}.
\newblock \bibinfo{journal}{\emph{IEEE Transactions on Parallel and Distributed
  Systems}} \bibinfo{volume}{33}, \bibinfo{number}{11} (\bibinfo{year}{2022}),
  \bibinfo{pages}{3220--3232}.
\newblock
\showISSN{1558-2183}


\end{thebibliography}

\end{document}